\documentclass[conference]{IEEEtran}
\IEEEoverridecommandlockouts

\usepackage{tikz}
\usepackage{amsmath}

\usepackage{fancyhdr}
\usepackage[normalem]{ulem}
\usepackage[hyphens]{url}
\usepackage{xspace}
\usepackage{booktabs}
\usepackage{graphicx}
\usepackage{makecell}
\usepackage{listings}
\usepackage{float}
\usepackage{placeins}
\usepackage{color,enumitem}
\usepackage{xcolor}
\usepackage{listings}
\usepackage{fontawesome}
\usepackage{blindtext}
\usepackage{makecell}
\usepackage{color,soul}
\usepackage{mathabx}
\usepackage{multirow}

\usepackage{balance}

\usepackage{afterpage}
\usepackage{xcolor}

\usepackage{cleveref}

\lstdefinestyle{cstyle}{
  belowcaptionskip=1\baselineskip,
  breaklines=true,
  xleftmargin=0em,
  language=C,
  tabsize=2,
  captionpos=t,
  showstringspaces=false,
  basicstyle=\scriptsize\ttfamily,
  keywordstyle=\bfseries\color{green!40!black},
  commentstyle=\itshape\color{purple!40!black},
  identifierstyle=\color{black},
  stringstyle=\color{orange},
  frame=tb,
}

\lstdefinestyle{asmstyle}{
  belowcaptionskip=1\baselineskip,
  xleftmargin=0em,
  language=[x86masm]Assembler,
  basicstyle=\footnotesize\ttfamily,
  commentstyle=\itshape\color{purple!10!black},
  keywordstyle=\bfseries\color{green!40!black},
  deletendkeywords={offset},
  morekeywords={p.bsw},
}

\newcommand{\mypar}[1]{\vspace{0.3em}\noindent\textbf{#1.}\hspace{0.5em}}
\newcommand{\mylist}[1]{\vspace{0em}\noindent\textbf{#1.}\hspace{0em}}

\newcommand{\name}{PsPIN\xspace}

\newcommand{\review}[1]{\hl{#1}}

\newcommand{\iscasubmissionnumber}{1399}
\title{A RISC-V in-network accelerator for flexible high-performance low-power  packet processing} 
\fancypagestyle{firstpage}{
	\fancyhf{}
	
	\fancyhead[C]{\normalsize{ISCA 2021 Submission
			\textbf{\#\iscasubmissionnumber} \\ Confidential Draft: DO NOT DISTRIBUTE}} 
	\fancyfoot[C]{\thepage}
}  

\usepackage{xparse}
\usepackage{marginnote}
\begin{document}

\DeclareDocumentCommand\review{m g g}{%
	{\IfNoValueF {#2}{%
			\IfNoValueF {#3}{%
				{\marginnote{\sethlcolor{#3}\hl{\normalfont \textbf{{\normalsize{\color{white}#2}}}}}%
				}%
			}%
			\IfNoValueT {#3}{%
				{\marginnote{\normalfont \textbf{\normalsize{#2}}}%
				}%
			}%
		}%
		\hl{#1}%
	}%
}

\DeclareDocumentCommand\review{m g g}{#1}

\author{
	\IEEEauthorblockN{Salvatore Di Girolamo\IEEEauthorrefmark{1}, Andreas Kurth\IEEEauthorrefmark{2},
		Alexandru Calotoiu\IEEEauthorrefmark{1}, Thomas Benz\IEEEauthorrefmark{2}, Timo Schneider\IEEEauthorrefmark{1}, \\
        Jakub Beránek\IEEEauthorrefmark{3}, Luca Benini\IEEEauthorrefmark{2}, Torsten Hoefler\IEEEauthorrefmark{1}}
    \vspace{0.8em}
	\IEEEauthorblockA{\IEEEauthorrefmark{1}Dept. of Computer Science, ETH Zürich, Switzerland\\}
	\IEEEauthorblockA{\IEEEauthorrefmark{2}Integrated System Laboratory, ETH Zürich, Switzerland\\}
	\IEEEauthorblockA{\IEEEauthorrefmark{3}IT4Innovations, VŠB - Technical University of Ostrava\vspace{0.8em}\\}

	{\rm \IEEEauthorrefmark{1}\{first.lastname\}@inf.ethz.ch, \IEEEauthorrefmark{2}\{first.lastname\}@iis.ee.ethz.ch, \IEEEauthorrefmark{3}jakub.beranek@vsb.cz}\vspace{-0.5em}\\
}
\date{}
\maketitle
\thispagestyle{empty}

%%%%%% -- PAPER CONTENT STARTS-- %%%%%%%%
\begin{abstract}
The capacity of offloading data and control tasks to the network
is becoming increasingly important, especially if we consider the faster growth
of network speed when compared to CPU frequencies. In-network compute alleviates the host CPU load by running tasks directly in the
network, enabling additional computation/communication overlap and potentially
improving overall application performance. 
However, sustaining bandwidths provided by next-generation networks, e.g., 400 Gbit/s, can become a challenge. sPIN is a programming model for
in-NIC compute, where users specify handler functions that are executed
on the NIC, for each incoming packet belonging to a given message or flow. It enables a CUDA-like acceleration, where the NIC is equipped with lightweight processing elements that process network packets in parallel. 
We investigate the architectural
specialties that a sPIN NIC should provide to enable high-performance, low-power, and flexible packet processing.
We introduce PsPIN, a first open-source sPIN implementation, based on a multi-cluster RISC-V architecture and designed according to the identified
architectural specialties.
We investigate the performance of PsPIN with cycle-accurate simulations, showing that it can process packets at 400 
Gbit/s for several use cases, introducing minimal latencies (26 ns for 64 B packets) and occupying a total area of 18.5 mm$^2$ (22 nm FDSOI).

%Network speed increases faster than CPU frequency, leading 

%- There is a growing difference between the evolution of networks speed and CPUs
%  frequency (CPU frequency grows slower than network speed). 

%- To counter-balance this gap, in-network compute is becoming increasingly
%  important. 

% 2. Specific Background: Zoom in from the thing everyone cares about to the
% thing you did.
%- In this context, sPIN is a programming model that has been introduced in
%  order to enable in-network compute. 

%- sPIN extends the NIC architecture to enable the users to run code, on a
%  per-packet basis, directly on the NIC. 

% 3. Statement of Problem or Knowledge Gap: What specific problem or phenomenon
% do we not understand in this field of study?

%- However, what are the architectural charactersitics a NIC should have
%  in order to support line-rate user-defined packet processing?

% 4. Here we show: One sentence about how you met the demonstrated need.
%- In this work, we define the architectural requirements of a
%  packet-processing-enabled NIC, taking sPIN as reference programming model. 

%- We also introduce P-sPIN, a sPIN unit prototype that can be integrated in
%  existing NIC architectures. 

% 5. Results: Only the very highest-level results. Sometimes a summary of
% approach comes before the summary of results.

\end{abstract}

\section{Motivation}

% packet processing becomes more important
Today's cloud and high-performance datacenters form a crucial pillar of
compute infrastructures and are growing at unprecedented speeds. At the core,
they are a collection of machines connected by a fast network
carrying petabits per second of internal and external traffic. Emerging online
services such as video communication, streaming, and online collaboration
increase the incoming and outgoing traffic volume. Furthermore, the growing
deployment of specialized accelerators and general trends towards
disaggregation exacerbates the quickly growing network load. Packet processing
capabilities are a top performance target for datacenters.
 
% we have RDMA today - slowly picks up speed
These requirements have led to a wave of modernization in datacenter networks: not
only are high-bandwidth technologies going up to 200 Gbit/s gaining wide
adoption but endpoints must also be tuned to reduce packet processing
overheads. Specifically, remote direct memory access (RDMA) networks move much
of the packet and protocol processing to fixed-function hardware units in the
network card and directly access data into user-space memory. Even
though this greatly reduces packet processing overheads on the CPU, the
incoming data must still be processed. A flurry of specialized technologies
exists to move additional parts of this processing into network cards, e.g.,
FPGAs virtualization support~\cite{firestone2018azure}, P4 simple rewriting rules~\cite{bosshart2014p4}, or triggered operations~\cite{portals42}.

\begin{table*}[h]
	\footnotesize
	\setlength{\tabcolsep}{5pt}
	\begin{tabular}{lccccp{10.6cm}}
		\textbf{Solution} & \textbf{L} & \textbf{P} & \textbf{G} & \textbf{U} & \textbf{Notes} \\
		\toprule
		Azure AccellNet~\cite{firestone2018azure} & \faThumbsUp & \faThumbsODown & \faThumbsOUp & \faThumbsODown & FPGA-based NICs; Flow-steering.   \\
		P4~\cite{bosshart2014p4}$^*$, FlexNIC~\cite{flexnic} & \faThumbsUp & \faThumbsOUp & \faThumbsOUp & \faThumbsODown & Packet steering and rewriting. FlexNIC adds memory support. \review{$^*$Runs on NICs and switches.}{B.3} \\
		Mellanox SHARP~\cite{graham2016scalable} & \faThumbsUp & \faThumbsODown & \faThumbsOUp & \faThumbsUp & Data aggregation and reduction. \review{Runs on switches.} \\
		Portals 4~\cite{portals42}, INCA~\cite{inca} & \faThumbsOUp & \faThumbsOUp & \faThumbsODown & \faThumbsUp & Sequences of predefined actions can be expressed with triggered operations. \review{Both target NICs.} \\
		Mellanox CORE-Direct~\cite{connectx} & \faThumbsUp & \faThumbsOUp & \faThumbsODown & \faThumbsUp & Sequence of predefined actions can be chained. \review{Targets switches.}\\
		Cray Aries Reduction Engine~\cite{alverson2012cray} & \faThumbsUp & \faThumbsODown & \faThumbsOUp & \faThumbsUp & Data reductions (up to 64 bytes). \review{Runs on switches.} \\
		Quadrics~\cite{petrini2001hardware}, Myrinet~\cite{buntinas2000fast} & \faThumbsUp & \faThumbsUp & \faThumbsOUp & \faThumbsODown & Users define threads to run on the NIC / NIC is re-programmable by users. \\
		SmartNICs~\cite{bluefield, broadcom} & \faThumbsOUp & \faThumbsUp & \faThumbsUp & \faThumbsODown & Runs full linux stack on the NIC; Offloading of new code requires flashing. \\
	    FPGA packet parsing pipeline~\mbox{\cite{attig2011400}} & \faThumbsUp & \faThumbsOUp & \faThumbsOUp & \faThumbsODown & Only packet parsing, read-only packets, might require FPGA reconfiguration. \review{Target NICs.} \\
		eBPF (host)~\cite{miano2018creating} & \faThumbsODown & \faThumbsOUp & \faThumbsOUp & \faThumbsODown &  Runs user-defined code (eBPF code) in virtual machine in the OS kernel. \\
		eBPF (Netronome)~\cite{ebpf-netronome}, hXDP~\cite{brunella2020hxdp} & \faThumbsUp & \faThumbsOUp & \faThumbsOUp & \faThumbsODown & eBPF programs can be offloaded to NIC. \\
		DPDK~\cite{dpdk} & \faThumbsODown & \faThumbsUp & \faThumbsOUp & \faThumbsODown & Runs in user space. Applications can poll for new raw packets from the NIC. \\
		StRoM~\mbox{\cite{sidler2020strom}} & \faThumbsUp & \faThumbsUp & \faThumbsODown & \faThumbsODown & Handlers for DMA streams are implemented on FPGA NIC. \\  
		NICA~\mbox{\cite{eran2019nica}} & \faThumbsUp & \faThumbsUp & \faThumbsODown & \faThumbsODown & Bind kernels running on on-NIC accelerators  to user sockets. \\
		\review{PANIC}~\cite{lin2020panic} & \faThumbsUp & \faThumbsUp & \faThumbsUp & \faThumbsODown & Applications compose execution of pre-installed compute units. Targets NICs.\\
		\textbf{sPIN}~\cite{spin} & \faThumbsUp & \faThumbsUp & \faThumbsUp & \faThumbsUp & Applications define C/C++ packet handlers to map to different messages/flows. \review{Targets NICs.}\\
		\bottomrule
	\end{tabular}
	\vspace{0.4em}
	\caption{
		\textbf{L}: Location (\faThumbsUp~on NICs or switches; \faThumbsOUp~on NICs but outside the packet pipeline; \faThumbsODown~on host CPU). \textbf{P}: Programmability
		(\faThumbsUp~fully programmable; \faThumbsOUp~limited programmability;
		\faThumbsODown~predefined functions). \textbf{G}: Granularity
		(\faThumbsUp~message and packets; \faThumbsOUp~only packets;
		\faThumbsODown~only messages). \textbf{U}: Usability (\faThumbsUp~usable by
		applications and privileged users; \faThumbsODown~only by privileged users). }
	\label{table:survey}
	\vspace{-2.5em}
\end{table*}

% spin rules them all
Streaming processing in the network (sPIN)~\cite{spin} defines a
unified programming model and architecture for network acceleration beyond
simple RDMA. It provides a user-level interface, similar to CUDA for compute acceleration, considering the specialties and constraints of low-latency line-rate packet processing. It defines a flexible and programmable network instruction set architecture (NISA) that not only lowers the barrier of entry but also supports a large set of use-cases~\cite{spin}.
For example, Di Girolamo et al.\ demonstrate up to 10x speedups for
serialization and deserialization (marshalling) of non-consecutive
data~\cite{spin-ddt}.

% how to implement the sPIN NISA?
While the NISA defined by sPIN can be implemented on existing
SmartNICs~\cite{broadcom}, their microarchitecture (often standard ARM SoCs) is not optimized for packet-processing tasks. In this work, we define an open-source high-performance and low-power microarchitecture for sPIN network interface cards (NICs). We break first ground by developing principles for NIC microarchitectures that enable flexible packet processing at 400 Gbit/s line-rate.

% core contributions
As core contributions in this work, we 
%\begin{itemize}[noitemsep,topsep=0pt,parsep=0pt,partopsep=0pt,leftmargin=*]
\begin{itemize}[noitemsep,topsep=0pt,parsep=3pt,partopsep=0pt,leftmargin=12pt]
    \item establish principles for flexible and programmable NIC-based packet processing microarchitectures,
    \item design and implement a fully-functional 32-core SoC for packet processing that can be added into any NIC pipeline,
    \item analyze latencies, message rates, and bandwidths for a large set of example processing handlers, and 
    \item open-source the SoC design to benefit the community.
\end{itemize}
\vspace{2pt}
We implement \name in synthesizable hardware description language (HDL) code. Overall, it occupies less 
than 20mm$^2$ in a 22nm FDSOI process, which is about 25x smaller than an Intel Skylake Xeon die. 
We show how, given the nature of in-network compute tasks, a PsPIN unit can achieve similar or higher
throughput than more complex architectures, like Xeon- or ARM- based ones by using at most 6.3W.

%\enlargethispage{0.5\baselineskip}

%\vspace{-1em}
\section{In-network compute}\label{sec:in-network-compute}\vspace{-0.4em}
%This is the theory part of the paper. We define the concept of in-network
%compute, discuss examples of how this can accelearte applications, and
%model the potential performance improvements. We can also make this section
%like a small survey and describe the different approaches to in-network
%compute (e.g., FPGA, ASIC, SmartNICs).

% What is network compute?
\emph{In-network compute} is the capability of an interconnection network to
process, steer, and produce data according to a set of programmable actions.
The exact definition of \emph{action} depends on the specific in-network-compute
solution: it can vary from pre-defined actions (e.g., pass or drop a packet
according to a set of rules) to fully programmable packet or message handlers
(e.g., sPIN handlers).
%
%The policies can be applied at different network levels (e.g., NICs or
%switches), they can be defined on full messages or single packets, they can be
%installed by user applications or by the operating system, and normally their
%execution does not involve the host CPUs. 

% Why is it important?
There are several advantages of computing in the network:
(1)~\emph{More overlap}. Applications can define 
actions to execute on incoming data.  Letting the network execute them
allows applications to overlap these tasks with other useful work; 
(2)~\emph{Lower latency}. The network can promptly react to incoming data (cf. Portals 4 triggered operations~\cite{portals42}, virtual functions~\cite{firestone2018azure}, sPIN handlers), immediately executing actions depending on it. Doing the same on the host requires applications to poll for new data, check for dependent actions, and then execute them.
(3)~\emph{Higher throughput}. Some in-network-compute solutions enable stream
processing of the incoming data. For example, sPIN can run packet handlers on
each incoming packet, potentially improving the overall throughput.  
(4)~\emph{Less resource contention}.  Running tasks in the network can
reduce the volume of data moved through the PCIe bus and the
memory hierarchy.  This implies fewer data
movements, less memory contention and cache pollution, potentially
improving the performance of host CPU tasks.

%\enlargethispage{0.5\baselineskip}

% How do we classify in-network compute solutions?
%In-network compute solutions range from specialized interfaces that allow
%applications to define rules for steering and/or rewriting
%packets~\cite{bosshart2014p4} to fully programmable NICs where the applications
%can offload data processing functions~\cite{spin}.  

Table~\ref{table:survey} surveys existing in-network-compute
solutions.
This classification focuses on the high-level characteristics of these solutions, comparing
them by the location where policies are run, the level of programmability, the granularity at which the actions are applied, and their usability. 

%While this classification does not distinguish between
%programming languages (e.g., P4), libraries (e.g., Portals 4), and architectures (e.g.,
%Azure AccellNet), it focuses on the high-level characteristics of these solutions, comparing
%them by the location where policies are run, the level of programmability, the granularity at which the actions are applied, and their usability. 

\mylist{(L)~Location} Policies can be executed at different points along the 
path from the endpoint sending the data to the endpoint receiving it. We
classify in-network-compute solutions as: \faThumbsUp~running in network
devices (e.g., on NICs or switches); \faThumbsOUp~running in network devices
but not on the packet pipeline (e.g., SmartNICs act as close-to-network endpoints, 
running full Linux stack); 
\faThumbsODown~if they run on the host CPUs.
%\enlargethispage{\baselineskip}

%\enlargethispage{\baselineskip}
\mylist{(P)~Programmability} It defines the expressiveness of the actions.
Network solutions enabling fully programmable actions that can access the
message/packet header and payload, access the NIC and host memory, and issue
new network operations (e.g., RDMA put or gets) are marked with~\faThumbsUp.
Solutions that provide a predefined set of actions that can be composed among
themselves (e.g., P4 match-actions or Portals 4 triggered operations) are
marked with~\faThumbsOUp. Solutions providing only predefined functions are
marked with~\faThumbsODown.

\mylist{(G)~Granularity} Actions can be applied to full messages
(\faThumbsODown), requiring to first fully receive the message, or to single
packets, as they are received (\faThumbsOUp).  Solutions enabling both types of
actions are marked with~\faThumbsUp.

\mylist{(U)~Usability} It defines which entities can install actions into
the network. In-network-compute solutions enabling user applications and
libraries (even in multi-tenant settings) to install actions are marked
with~\faThumbsUp.  Solutions that require elevated privileges, service
disruption, and/or device memory flashing to install new actions are marked
with~\faThumbsODown.
\vspace{0.4em}

%\enlargethispage{0.5\baselineskip}

Among the solutions of Table~{\ref{table:survey}}, sPIN is the only one letting
user-space applications define per-message or per-packet tasks (called handlers) that are executed in the
NIC. sPIN handlers can access and modify packet data,
share NIC memory, and issue NIC and DMA commands. Handlers can be installed on the 
NIC without disrupting operations and memory isolation must be guaranteed (see Section~{\ref{sec:intra-sched}}).
For these reasons, we focus on the sPIN programming model, investigate the 
challenges of building a sPIN engine, and introduce \name, a general and open-source 
sPIN implementation.
\review{By open-sourcing the hardware design of \name{} under a permissive open-source license (Solderpad), we want to encourage its usage and foster the creation of prototypes by anyone in the community.}{A.2}

%that runs in the network
%(specifically in NICs) and lets the users express per-message or per-packet
%functions (defined in C or C++, called handlers) from which they can access
%5packet data, share NIC memory, and issue NIC and DMA commands.  Moreover, the
%handlers can be defined by user applications and do not require
%disruptions of the NIC operation.  For these reasons, this work focuses on
%the sPIN programming model, investigating the challenges of building a
%sPIN engine, and introducing \name, a general and open-source sPIN implementation that can be integrated into any NIC design.

%\vspace{-1em}
%\enlargethispage{\baselineskip}
\subsection{sPIN: Streaming processing in the network}
\vspace{-0.3em}

\begin{figure}[t]
	\centering
	\includegraphics[width=\columnwidth]{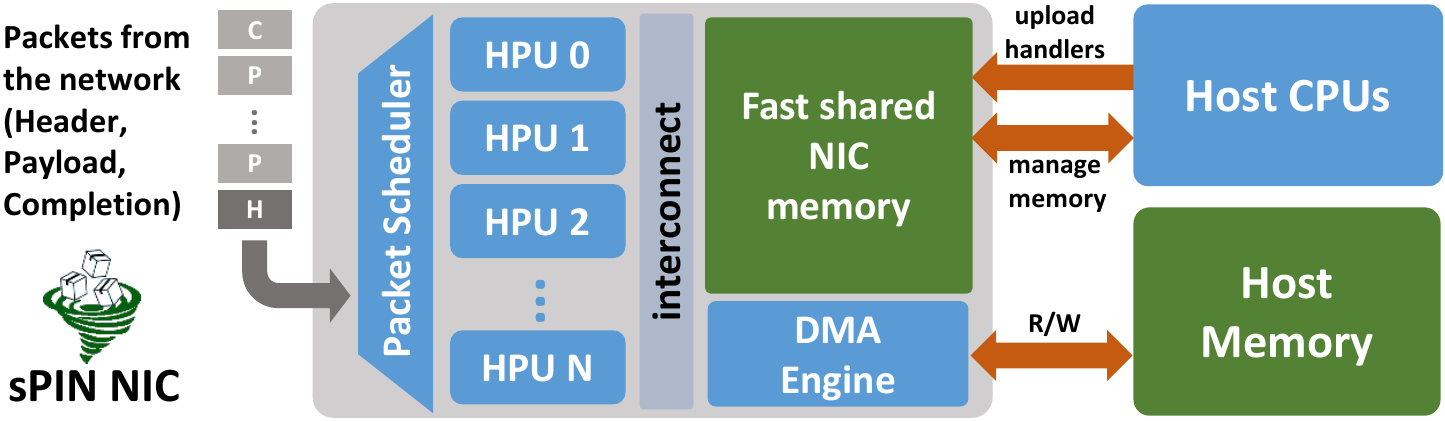}
	\vspace{-2.1em}
	\caption{sPIN abstract machine model.}
	\vspace{-1.5em}
	\label{fig:spin-abm}
\end{figure}

The key idea of sPIN is to extend RDMA by enabling users to define simple processing tasks, called \emph{handlers}, to be executed directly on the NIC. 
A message sent through the network is seen as a sequence of packets: the
first packet is defined as \emph{header}, the last one as \emph{completion},
and all the intermediate ones as \emph{payload}. 
As the packets of a message reach their destination, the receiving NIC invokes the
respective packet handlers. For each message, three types
of handlers are defined: the \emph{header handler}, executed only on the header
packet; the \emph{payload handler}, executed on all the packets, and the \emph{completion handler}, executed after all packets have been processed.
Handlers are defined by applications running on the host and cross-compiled for the NIC microarchitecture. The programming model that sPIN proposes is
similar to CUDA~\cite{nickolls2008scalable} and
OpenCL~\cite{stone2010opencl}: the difference is that in these frameworks,
applications define kernels to be offloaded to GPUs. In sPIN, the kernels
(i.e., handlers) are offloaded to the NIC, and their execution is triggered by
the arrival of packets. Figure~\ref{fig:spin-abm} sketches the sPIN abstract
machine model.

Host applications define packet handlers and associate them with message descriptors. Packet handlers are optional:
e.g., by specifying either the header or completion handler and no payload handler, only one packet handler
for the full message will be executed. Message descriptors, together with
packet handlers, are installed into the NIC. 
Incoming packets are matched to message descriptors and handlers are 
executed on Handler Processing Units (HPUs).
Handlers can also issue NIC commands and DMA transfers to/from the host memory.

%\vspace{-1em}
\subsubsection{Architectural Specialties}\label{sec:requirements}
%\vspace{-0.5em}

%\enlargethispage{0.5\baselineskip}

The sPIN abstract machine model specifies a streaming execution model with microarchitectural requirements that are quite different from classical specialized packet processing engines, which normally constraint the type of actions that can be performed or the entity that can program them, and traditional compute cores.  We now outline a set of architectural properties that a sPIN implementation should provide to enable fully-programmable high-performance packet processing.

\mylist{S1.~Highly parallel} Many payload packets can be processed in parallel. The higher the number of HPUs, the longer the handlers can run without becoming a bottleneck.

\mylist{S2.~Fast scheduling} Arriving packets must be scheduled to HPU cores while maintaining ordering requirements that mandate that header handlers execute before payload handlers that execute before completion handlers. 

\mylist{S3.~Fast explicit memory access} Packet processing has low temporal locality by definition (a packet is seen only once), hence scratchpad memories are better than caches. 

\mylist{S4.~Local handler state} Handlers can keep state across packets of a
message as well as multiple messages. If the memory is partitioned, then
scheduling needs to ensure that the state is reachable/addressable.
 
\mylist{S5.~Low latency, full throughput} To minimize the time a packet stays in
the NIC, the time from when the packet is seen by sPIN to when the handlers
execute should be minimized. Furthermore, the sPIN unit must not obstruct
line-rate.

\mylist{S6.~Area and power efficiency} To lead to an
easier integration of a sPIN unit in a broader range of NIC architectures.

%The sPIN unit should be small enough to be
%integrated in existing NIC architectures. For reference, Mellanox Bluefield
%SmartNIC is equipped with 16 A72 ARM cores (64 bit) with 1 MiB shared L2 caches
%per 2 cores, and 6 MiB shared L3 last-level cache. The total area needed by
%this computing unit (cores + caches) is about 5.66 mm$^2$, assuming 22nm
%technology. A sPIN unit should not exceed this area.  \salvo{add more examples?
%cite}

\mylist{S7. Handler isolation} Handlers processing a message should not be able
to access memory belonging to other messages, especially if they belong to
different applications.

\mylist{S8. Configurability} A sPIN unit should be easily re-configurable to be
scaled to different network requirements. 

%\mypar{S8. Virtualization support} A sPIN unit should provide a clear interface
%for migrating the message processing to a different NIC in case of
%virtual-machine migration.

%\vspace{-1em}
\section{\name}
\vspace{-0.4em}

%\enlargethispage{0.5\baselineskip}

%\begin{figure*}[t]
%    \centering
%    \includegraphics[width=\textwidth]{img/pspin-crop.pdf}
%    \caption{\name overview. the figure shows the NIC model (on the left), with
%    a zoom on the \name unit.}
%    \label{fig:pspin-overview}
%\end{figure*}

\name is a sPIN implementation designed to match the architecture specialties of Section~\ref{sec:requirements}.
PsPIN builds on top of the PULP (parallel ultra-low power) platform~\cite{pulp}, a silicon-proven~\cite{gautschi2017near} and open~\cite{traber2016pulpino} architectural template for scalable and energy-efficient processing.  PULP implements the RISC-V ISA~\cite{waterman2014risc} and organizes the processing elements in
clusters: each cluster has a fixed number of cores (32-bit, single-issue, in-order) and single-cycle-accessible scratchpad memory (\textbf{S3}). The system can be scaled by adding or removing clusters (\textbf{S1}).
\emph{We have implemented all hardware components of \name in synthesizable hardware description language (HDL) code.}

%We now describe the \name architecture (\cref{sec:architecture}), then
%discuss how control- and data- plane operations are implemented in \name. 
%(\cref{sec:runtime}). Finally, we discuss how this packet-processing unit can
%be integrated in existing NIC architectures (\cref{sec:integration}).

%\enlargethispage{0.5\baselineskip}
%\vspace{-1em}
\subsection{Architecture Overview}\label{sec:architecture}
\vspace{-0.3em}

\begin{figure}[h]
	\vspace{-0.8em}
	\centering
	\includegraphics[width=\columnwidth]{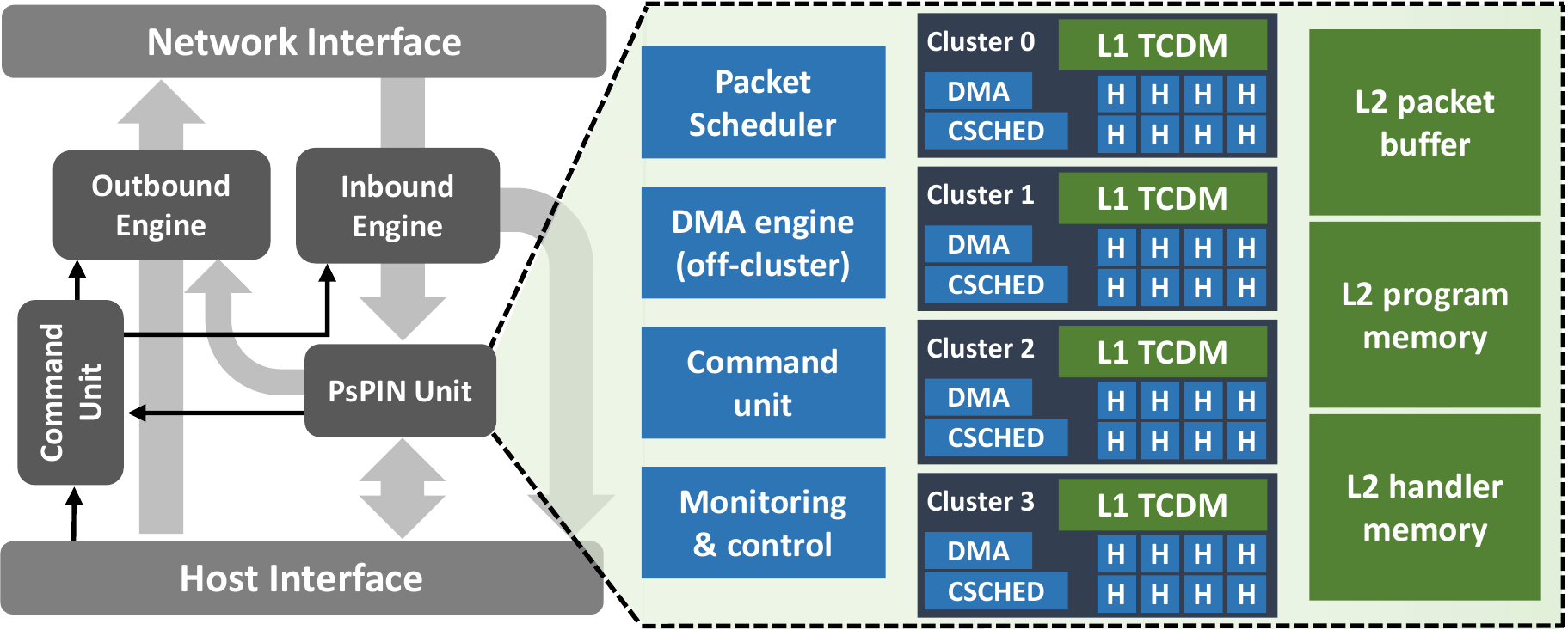}
	\vspace{-2em}
	\caption{NIC model and \name architecture overview.}
	\vspace{-0.5em}
	\label{fig:arch-overview}
\end{figure}

\name has a modular architecture, where the HPUs are
grouped into processing clusters. The HPUs are implemented as
\mbox{RISC-V} cores, and each cluster is equipped with a 
single-cycle access scratchpad memory called L1 memory.
All clusters are interconnected to each other (i.e., HPUs can
access data in remote L1s) and to three off-cluster memories (L2): 
the packet buffer, the handler memory, and the program memory. 
Figure~\ref{fig:arch-overview} shows an overview of how \name integrates
in a generic NIC model and its architecture. 
We adopt a generic NIC model to identify the general building blocks of a NIC architecture. Later, in Section~\ref{sec:integration}, we discuss how \name can be integrated in existing NIC architectures.

%The host applications can allocate handler memory through the NIC driver and can
%write or read from it. This memory is used to let the host copy data that
%is necessary to the handlers (e.g., packet filtering rules). 
%The handlers code is copied by the host to program memory. 
%The host has no access to the packet buffer, which is accessed by the
%NIC inbound engine to write packets to process, by the processing elements and the %clusters' DMA engines. 

Host applications access program and handler memories
to offload handlers code and data, respectively. The management
of these memory regions is left to the NIC driver, which is in charge
of exposing an interface to the applications in order to move code and data.
The toolchain and the NIC driver extensions to offload handlers code and data are out of the scope of this work. 
Once both code and data for the handlers are offloaded, the host builds
an execution context, which contains: pointers to the handler
functions (header, payload, and completion handlers), a pointer to the
allocated handler memory, and information on how to match packets
that need to be processed according to this execution context. The execution
context is offloaded to the NIC and it is used by the NIC inbound engine
to forward packets to \name.

%
%Since \name is not tied to a particular NIC architecture, we 
%do not specify the details of how packets
%are matched to execution contexts in this section. A discussion on  how
%the matching process can be either extend or added to existing NIC %architectures is included in Section~\ref{sec:integration}.

%\enlargethispage{0.5\baselineskip}

% need for matching
\mypar{Receiving data}
Data is received by the NIC inbound engine, which is normally
interfaced with the host for copying it to host memory. In a 
\name-NIC, the inbound engine is also interfaced to the \name unit.
The inbound engine must be able to distinguish packets that need to
be processed by \name from the ones taking the classical non-processing path.
To make this distinction, it matches packets to
\name execution contexts and, if a match is found, it forwards the packet to the \name unit. Otherwise, the packet is copied to the host as normal. While some networks already have the concept of packet matching (e.g., RDMA NICs match packets to queue pairs), in others 
this concept is missing and needs to be introduced to enable
packet-level processing (see Section~\ref{sec:integration}).  

%sending data to PsPIN
%\mypar{Handler Execution Request (HER)}
Packets to be processed on the NIC are copied to the
L2 packet buffer. Once the copy is 
complete, the NIC inbound sends a Handler Execution Request (HER) to 
\name's packet scheduler. The HER contains all information necessary to schedule a handler to process the packet, which are a pointer to the packet in the L2 packet buffer and an execution context. 
If the packet buffer is full, the NIC inbound engine can either back pressure the senders~\cite{infiniband2000infiniband}, send explicit congestion notifications~\cite{ramakrishnan2001addition}, drop packets, or kill connections~\cite{portals42}. The exact policy to adopt depends on the network in which \name
is integrated and the choice is similar to the case where the host cannot consume incoming packets fast enough.

The packet scheduler selects the processing cluster that processes
the new packet. The cluster-local scheduler (CSCHED) is in charge
of starting a DMA copy of the packets from the L2 packet buffer to the L1  Tightly-Coupled Data Memory (TCDM) and selecting an idle HPU (H) where to run handlers for packets that are available in L1. Once the packet processing completes, a notification is sent back to the NIC to let it update its view of the packet buffer (e.g., move the head pointer in case the packet buffer is managed as a ring buffer).

%getting data out of PsPIN
\mypar{Sending Data}
Packet handlers, in addition to processing the packet data, can send data
over the network or move data to/from host memory. To send data directly from the NIC, the
sPIN API provides an RDMA-\emph{put} operation: When a handler issues this
operation, the \name runtime translates it into a NIC command, which is sent to the NIC outbound engine. If the NIC outbound engine cannot receive new commands, the handler blocks waiting for it to become available again. The NIC outbound can send data from either the L2 packet memory, the L2 handler memory, or L1 memories, or it can specify a host memory address as data source, behaving as a host-issued command. 
To move data to/from the host, the handlers can issue DMA operations:
These operations translate to commands that are forwarded to the off-cluster DMA engine, which writes data to host memory through PCIe.

\definecolor{orange}{RGB}{197,90,17}
\newcommand*\negcircnum[1]{\tikz[baseline=(char.base)]{%
            \node[white,shape=circle,fill=orange,draw,inner sep=1pt] (char) {\color{white}\sffamily #1};}}

\vspace{-0.2em}
\subsection{Control path}
\vspace{-0.3em}

%\enlargethispage{0.5\baselineskip}

\begin{figure}[t]
	\centering
	\includegraphics[width=\columnwidth]{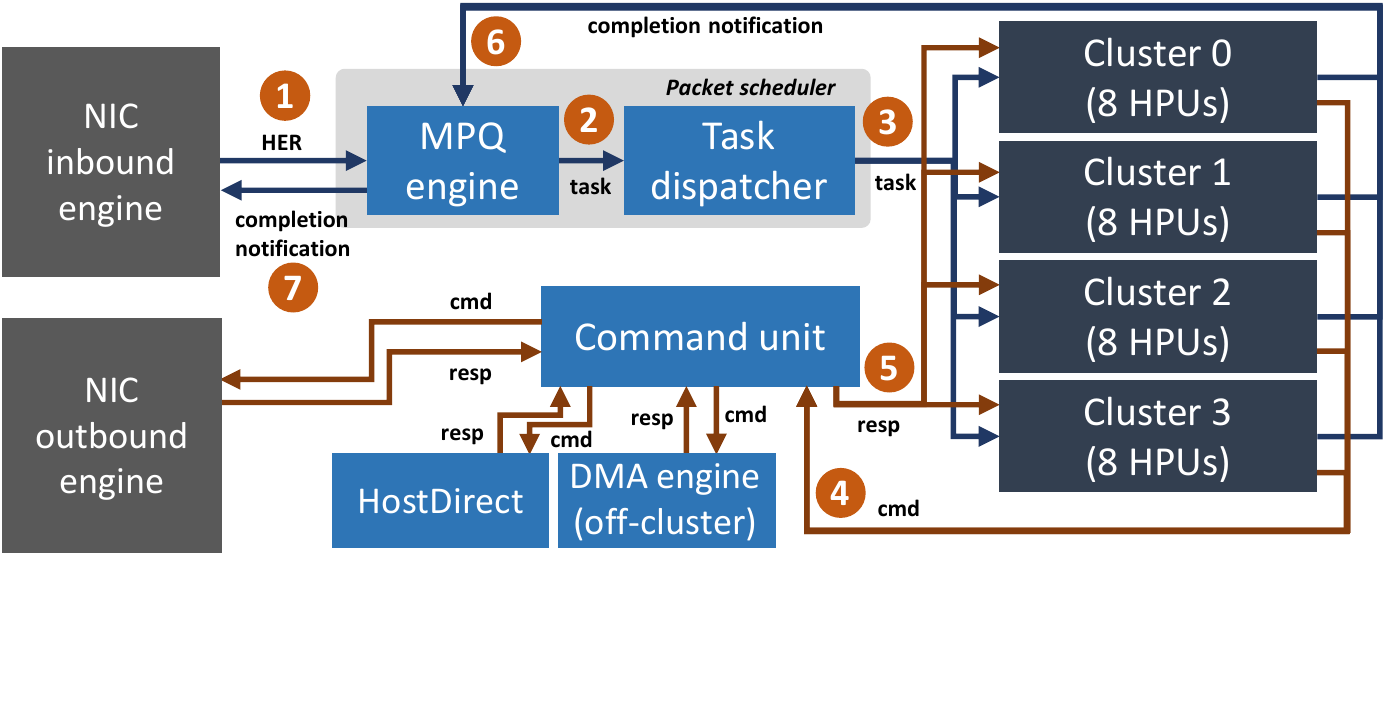}
	\vspace{-4.8em}
	\caption{PsPIN control path overview.}
	\vspace{-1.9em}
	\label{fig:control-path}
\end{figure}

Figure~\ref{fig:control-path} shows the \name control path, which includes:
\negcircnum{1}~receiving HERs from the NIC inbound engine, \negcircnum{2}~scheduling packets, handling
commands from the handlers, and \negcircnum{7}~sending completion notifications back to 
the NIC. 

\subsubsection{Inter-cluster packet scheduling}
%\vspace{-0.5em}
\name is informed of new packets to process by receiving
HERs from the NIC inbound engine~\negcircnum{1}. The HER is received by the packet scheduler, which is composed of the Message Processing Queue (MPQ) engine and the task dispatcher. The MPQ engine handles scheduling dependencies between packets. These dependencies are defined by the sPIN programming model:
\begin{itemize}[noitemsep,topsep=1pt,parsep=0pt,partopsep=0pt,leftmargin=*]
	\item the header handler is executed on the first packet of a message and no payload handler can start before its completion;
	\item the completion handler is executed after the last packet of a message is received and all payload handlers are completed.
\end{itemize}
A \emph{message} is a sequence of packets mapped to an MPQ and matched to an execution context. We let the NIC define the packets that are part of a message or flow. Once the last packet of a message arrives, the NIC marks the corresponding HER with an end-of-message flag, letting \name run the completion handler when all other handlers of that MPQ complete. 
%
%sPIN does not define inter-message dependencies but handlers of different messages can still synchronize by sharing memory and using atomic operations.

%To enforce scheduling dependencies, the MPQ engine organizes HERs in linked lists, one per message. If a packet is blocked, e.g., because the header handler is still running, its HER is queued in the linked list corresponding to its message. The MPQ engine then selects a ready message queue (i.e., no unsatisfied scheduling dependencies and not empty), from which to generate a processing task in a round-robin manner, and forwards it to the task dispatcher~\negcircnum{2}. 
%
%This approach allows us to have fair scheduling between messages in case different messages are received at the same time. 
%We choose to organize blocked HERs in linked lists because, under normal operations, a message is not in a blocked state and its packets should be scheduled at line rate. Hence, an approach with statically allocated FIFO buffers would result in a waste of memory cells. However, to avoid the case where a message consumes all the buffer space in the MPQ engine, we statically allocate four cells for each MPQ, allowing other messages to progress even in case a message blocks. 

\mypar{Task dispatcher} The task dispatcher selects the processing cluster where to forward a task for its execution~\negcircnum{3}. 
\review{We introduce a dedicated hardware unit for dispatching packets to clusters. A software solution would not provide enough bandwidth to schedule packets at line rate. If we consider a target bandwidth of 400 Gbit/s and 64 B packets, we get one packet every 1.28 ns, which requires to schedule a packet every 1.28 cycles on average. }{A.4}
A cluster can accept new tasks when it has enough space in its L1 to store the packet data.
%A task can be forwarded to a cluster if that cluster has enough space in its L1 to store the packet data.
%
We use the message ID, which is included in the HER, to determine the 
\emph{home cluster} of a message: the task dispatcher
tries to schedule packets to their home clusters. 
If the home cluster cannot accept it, then the 
least loaded cluster is selected. The task dispatcher blocks if there 
are no available clusters.

\begin{figure}[h]
	\vspace{-1.4em}
	\centering
	\includegraphics[width=\columnwidth]{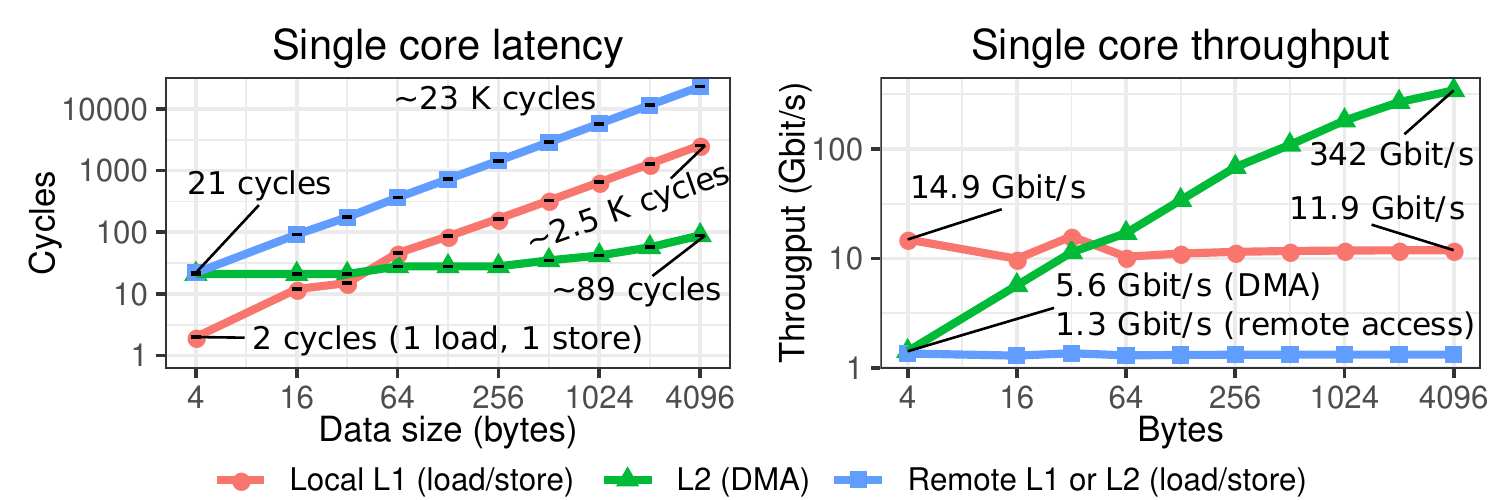}
	\vspace{-2em}
	\caption{\review{Data copy latency and bandwidth.}{E.3}}
	\vspace{-1em}
	\label{fig:mem-lat}
\end{figure}

The rationale behind the concept of home cluster is given by the fact
that handlers processing packets of the same message can share L1 memory, hence
scheduling them on the same cluster avoids remote L1 accesses. Figure~\ref{fig:mem-lat} shows the memory latency and bandwidth experienced by a single core when copying data from local or remote memories using different access types (i.e., load/stores, DMA).  As each core can execute one single-word memory access at a time, the latency for accessing a chunk of data increases linearly with its size.  The DMA engine, on the other hand, moves data in bursts, so multiple words can be ``in-flight'' concurrently.

%\enlargethispage{0.5\baselineskip}

\mypar{Handler execution and completion notification} 
Within a processing cluster, task execution requests are handled
by the cluster-local scheduler. We describe the details of intra-cluster
handler scheduling in Section~\ref{sec:intra-sched}. During their execution, handlers can issue commands that are handled by a command unit~\negcircnum{4}. 
We define three types of commands to interact with the NIC outbound and with the off-cluster DMA engine:
\begin{itemize}[noitemsep,topsep=0pt,parsep=3pt,partopsep=0pt,leftmargin=12pt]
 \item NIC commands to send data over the network: a handler can forward the packet or generate new ones. 
 \item DMA commands to move data to/from host memory. The host virtual addresses can be stored in application-defined data structures in handler memory. 
 \item HostDirect commands are similar to DMA commands but, instead of a source address, they carry 32 B immediate data that is written directly to the host memory address.
\end{itemize} 
Command responses~\negcircnum{5} are used to inform the handlers of the completion of the issued commands or error conditions.

Once a handler terminates and there are no in-flight commands for which
a response is still pending, a completion notification is generated~\negcircnum{6}. The MPQ engine uses this notification to track the state of message queues (e.g., mark a queue as ready when the header handler completes). The notification is also forwarded to the NIC inbound engine, which uses it to free sections in the L2 packet buffer. 

%\enlargethispage{0.5\baselineskip}

\begin{figure}[h]
	\vspace{-1em}
	\centering
	\includegraphics[width=\columnwidth]{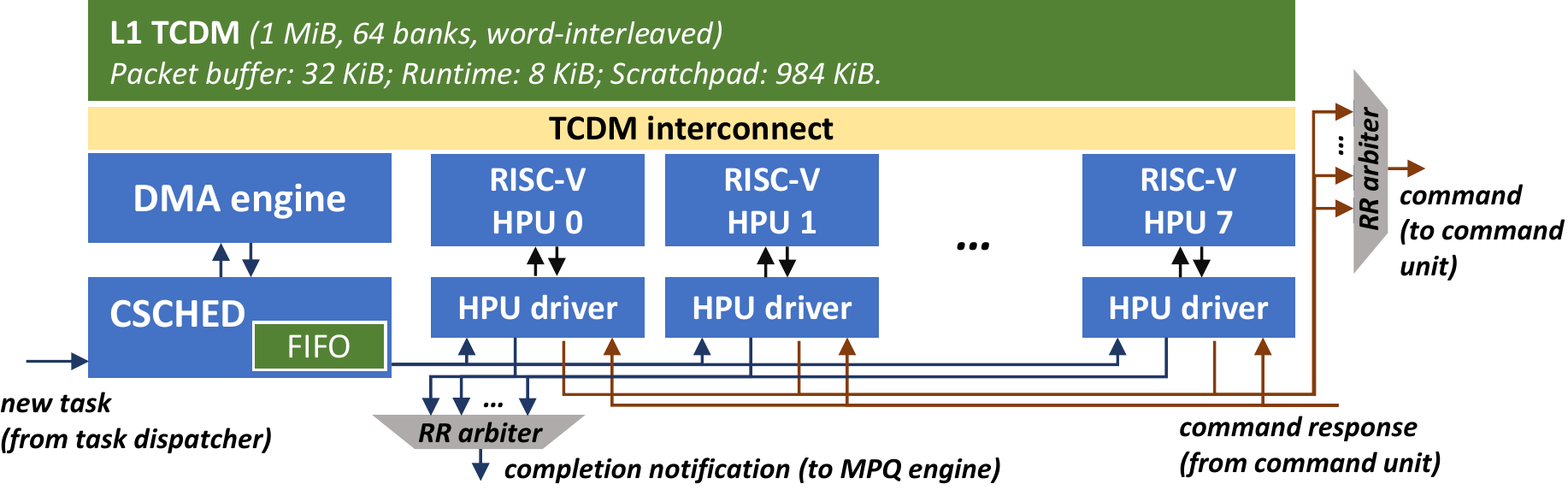}
	\vspace{-2em}
	\caption{\name processing cluster.}
	\vspace{-0.5em}
	\label{fig:cluster-detail}
\end{figure}

%\vspace{-1.5em}
\subsubsection{Intra-cluster handler scheduling}\label{sec:intra-sched}
%\vspace{-0.5em}

%While HPUs can access the packet data stored in the L2 packet buffer,
%memory accesses to L2 memories take up to 25 cycles (assuming no contention). To save this 
%latency, applications can specify in the execution context the number
%of bytes of the packet that must be made available in the L1 of the cluster
%where the handler is executing, enabling single-cycle access to this data.  
%This information is propagated into the task descriptor. 

Tasks are received by the cluster-local scheduler (CSCHED) that
is in charge of scheduling them on the HPUs. 
For each new task, it starts a DMA transfer of the packet data from L2 to L1. Moving the packet data to L1 enables 
single-cycle access from the cores.
%Tasks that
%are waiting for a DMA transfer to complete are buffered in a FIFO queue 
%(the DMA engine guarantees in-order completion of the transfers). 
Once a transfer completes, the corresponding task is popped from the queue and
scheduled to an idle HPU. 
{\reversemarginpar\review{At 400 Gbit/s, a cluster receives tasks for 64 B packets every 5.12 ns on average (1.28 ns $\cdot$ 4). 
This time budget is not sufficient to handle intra-cluster scheduling in software. For comparison, issuing a DMA command already takes 6 cycles. 
Furthermore, having the scheduling algorithm running on the HPUs, e.g., in a cooperative scheduling approach, would require to run it in a higher privilege mode in order to guarantee handlers isolation, adding additional overheads. Hence, we opted for a hardware intra-cluster scheduler, which also allowed us to have a lighter runtime running on the HPUs.}{A.4}}
HPUs are interfaced with a memory-mapped device, the HPU driver, from which they can read information about the task to execute. 

The \name runtime running on the HPU consists of a loop executing the following steps:
(1)~Read the handler function pointer from the HPU driver. If the HPU driver has no task/handler to execute, it stops the HPUs by clock-gating it. When a task arrives, the HPU is enabled and the load completes.
(2)~Prepare the handler arguments (e.g., packet memory pointer).
(3)~Call the handler function.
(4)~Write to a doorbell memory location in the HPU driver to inform it that the handler execution is completed.
The HPU driver sends a completion notification as soon as it detects that there are no in-flight commands issued by the completed task. The HPU driver can buffer a completed task for which the completion notification cannot be sent and start processing a new one.
%The new handler blocks if it issues a command while the HPU driver is still waiting for sending the notification of the previous handler.

Since multiple HPU drivers can send feedback and issue commands at the same time, 
we use round-robin arbiters to select, at every cycle, an HPU that can send a feedback and
one that can issue a command.
Figure~\ref{fig:cluster-detail} shows an overview of a \name processing
cluster. The figure shows only the connections relevant to
the scheduling processes and to the handling of handler commands. 
In reality, the HPUs are also interfaced to the cluster DMA engine and can
issue arbitrary DMA transfers from/to the accessible L2 handler memory. 

\mypar{Memory accesses and protection} 
Handlers processing packets matched to the same execution context share the L2 handler memory
region that has been allocated by the application when defining the execution context. Additionally,
each message shares a statically allocated scratchpad area in the L1 of the home cluster. In particular,
L1 memories, which are 1~MiB each in our configuration, contain: the packet buffer (32~KiB), 
the runtime data structure (e.g., HPU stacks, 8~KiB), the message scratchpads (984~KiB). 
Scratchpads are allocated through the NIC driver and associated with execution contexts.

%The size
%of the per-message scratchpad depends on the maximum number of messages that we allow to be in \name 
%at the same time. The current configuration allows for 512 in-flight messages, which are evenly
%distributed among the clusters (by the message ID), leading to a $\sim$7.6~KiB scratchpad per message.

To protect against illegal memory accesses and guarantee handler isolation (\textbf{S7}), the HPU driver
configures the RISC-V Physical Memory Protection (PMP) unit~\cite{waterman2016risc} for each task, 
allowing the core
to access only a subset of the address space (e.g., handler code, packet memory, L1 scratchpad). 
The handlers are always run in user mode. In case of a memory access violation or any other exception,
an interrupt is generated and handled by the \name runtime. The exception handling consists of
resetting the environment (e.g., stack pointer) for the next handler execution and informing
the HPU driver of the error condition. The HPU driver will then send a command to the HostDirect
unit to write the error condition to the execution context descriptor in host memory.  A failed handler leads to the release of the occupied resources. 

%\vspace{-0.5em}
\subsubsection{Monitoring and control}\label{sec:control}
%\vspace{-0.5em}
While processing packets on the NIC, there are two scenarios that must be prevented to ensure correct operation: (1)~Packets of
a message stop coming and the end-of-message is not received. This can be due to factors such as network failure, network congestion, or bugs in applications or protocols. (2)~Slow handlers that cannot process packets at line rate.
To detect case (1), we use a pseudo-LRU~\cite{handy1998cache} solution on active MPQs (i.e., MPQs which are receiving packets). Every time an MPQ is accessed (i.e., packet pushed to it), it is moved to the back of the LRU list. If the candidate victim does not receive packets for more than a threshold specified in the execution context of the message that activated it, the MPQ is reset. This event is signaled to the host through the execution context descriptor.
Case (2) is detected by the HPU drivers themselves by using a watchdog timer
that generates an interrupt on the HPU and causes the runtime to reset it. The timer is configured according to a threshold specified in the execution context either by the NIC driver or the application itself. This case is handled similarly to memory access violations by notifying the host of the error condition through the execution context descriptor.

\begin{figure}[t]
	\vspace{0em}
	\centering
	\includegraphics[width=\columnwidth]{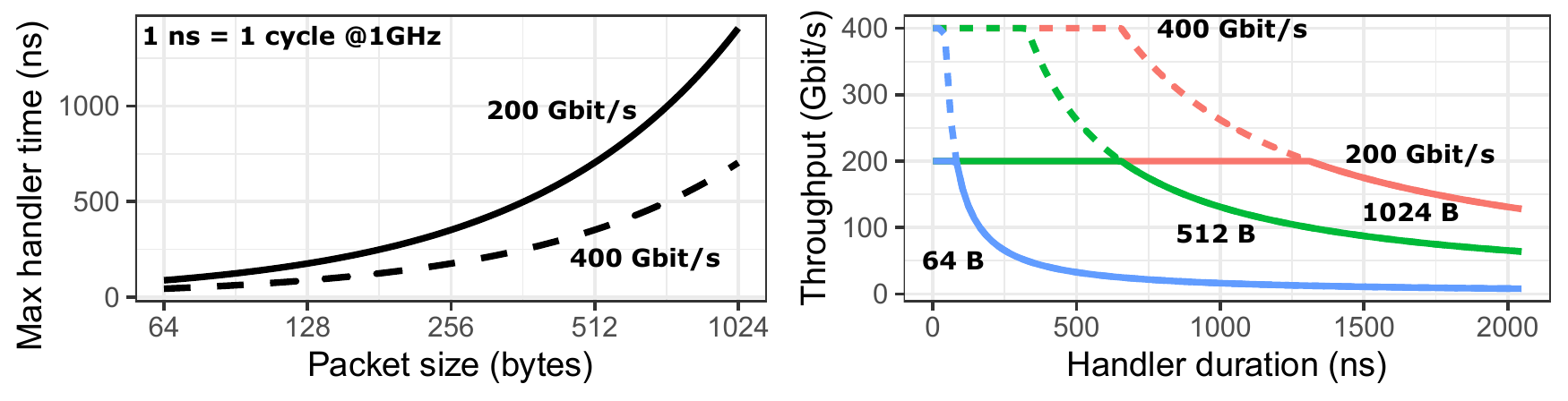}
	\vspace{-2.2em}
	\caption{Relation between handlers execution times and line rate.}
	\vspace{-1.5em}
	\label{fig:handler-model}
\end{figure}

To understand the time budget available to the handlers, Figure~\ref{fig:handler-model} shows the relation between handlers execution time and line rate. We assume a \name configuration with 32 HPUs. On the left, it shows the maximum duration handlers should
have to process packets at line rate for different packet sizes, in case of 200 Gbit/s and 400 Gbit/s networks. 
On the right, it shows how the processing throughput is affected by handlers
duration for different packet sizes and network speeds.

\vspace{-0.2em}
\subsection{Data path}\label{sec:datapath}
\vspace{-0.3em}

We now discuss how data flows within \name, explaining the design choices
made to guarantee optimal bandwidth. 
We equip \name with three interconnects: the NIC-Host interconnect, which interfaces the NIC and the host to \name memories; 
the DMA interconnect, which interfaces the cluster-local DMA engines to both L2 packet buffer and handler memories; and the processing-elements (PE) interconnect, which allows HPUs to read from either L2 memories or remote L1s. Both NIC-Host and DMA interconnect
have wide data ports (512 bit), while the PE interconnect is designed for finer 
granularity accesses (32 bit).  
%\salvo{why not 32? Because of the IC? Andreas: For legacy reasons (the old DMA had 64 bit data width, and the core-to-AXI module is still hardcoded to that value).  To avoid this explanation, we can say the PE interconnect is 32 bit, even though it is currently 64 bit -- this is only to our disadvantage (although it has very little impact on area and timing); see remark on throughput after next sentence.}
Since \name is clocked at 1 GHz, the offered bandwidth of these interconnects is  512 Gbit/s and 32 Gbit/s, respectively.
\name's on-chip interconnects, memory controllers, and DMA engine are based on~\cite{kurth2020axi}.
%\salvo{makes sense? Andreas: Yes.  Also, the PE interconnect effectively offers only 32 Gbit/s, because all transactions through it are 32 bit wide.}

\begin{figure}[h]
	\centering
	\includegraphics[width=\columnwidth]{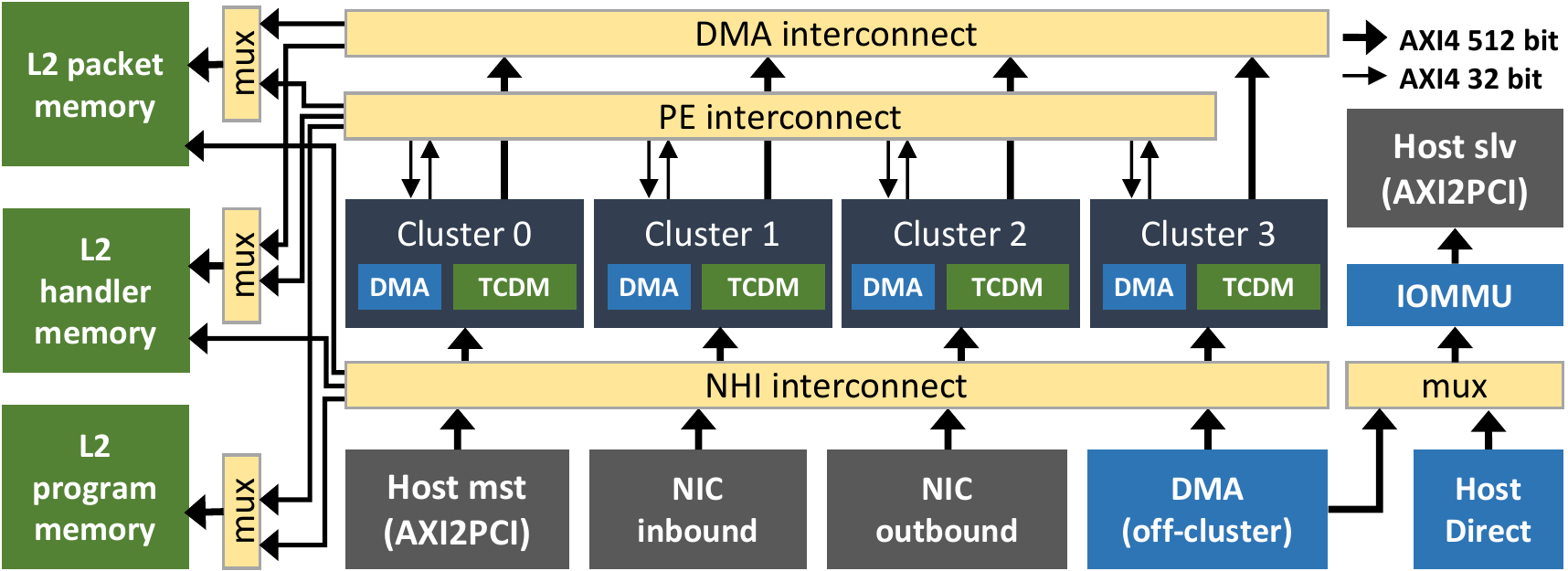}
	\vspace{-1.5em}
	\caption{\review{PsPIN data path overview. Bold arrows represent AXI4 connections with 512 bit data width. Thin arrows represent 32 bit AXI4 connections. Arrow heads indicate AXI4 ``slave'' ports, while tails are for ``master'' ports.}{E.3}}
	\vspace{-1.7em}
	\label{fig:data-path}
\end{figure}

Figure~\ref{fig:data-path} shows an overview of the \name memories, interconnects, and units that can move data (in gray if they are interfaced to but not within \name). We identify three critical data flows that require full bandwidth in order to not obstruct line rate and optimize \name data paths to achieve this goal.
\begin{itemize}[noitemsep,topsep=0pt,parsep=3pt,partopsep=0pt,leftmargin=12pt]
	\item \emph{Flow~1:~from NIC inbound to L2 packet buffer to clusters' L1s.} The NIC inbound writes packets to the L2 packet buffer at line rate and, in the worst case, this data is always copied to the L1s of the processing clusters by their DMA engines, before starting the handlers. The main bottleneck of this data flow can be the L2 packet buffer, which is accessed in both write and read directions. 
	
	\item \emph{Flow~2:~from L2/L1 to host memory.} Assuming all handlers copy the  data to host, we have a steady flow of data towards the host memory. The data source is specified in the command issued by the handlers and can be either the L2 packet buffer, the L2 handler memory, or the clusters' L1s. This data is moved by the off-cluster DMA engine, which interfaces to an IOMMU to translate the virtual
	addresses specified in the handler command to physical ones. The IOMMU is updated by the NIC driver when the host registers memory that can be accessed by the NIC. 
	
	\item \emph{Flow~3:~from L2/L1 to NIC outbound.} Similar to flow 2, but the data is moved towards the NIC outbound engine. We assume the NIC outbound has its own DMA engine, which it uses to read data.
	
\end{itemize}

All the identified critical flows can involve the L2 packet buffer. 
To avoid being a bottleneck, this memory 
must provide full bandwidth to the NIC inbound engine and to the cluster-local DMA
engines (flow 1), plus it must provide full bandwidth to the system composed of the
NIC outbound engine and the off-cluster DMA (flow 2 + flow 3), letting them reach up to 256 Gbit/s read-bandwidth each under full load. To achieve this goal, we
 implement the L2 packet buffer as 4 MiB, two-ports full-duplex, multi-banked (32 banks) word-interleaved memory. With 512 bit words, the L2 packet buffer is suitable more
for wide accesses than single (32 bit) load/store accesses from HPUs. In fact,
if handlers are going to frequently access packets, then their execution context
can be configured to let \name move packets to L1, before the handlers start. 
The maximum bandwidth that the L2 packet buffer can sustain is 512 Gbit/s per port, full duplex. This bandwidth can be achieved in case there are no bank conflicts. 
One port of the L2 packet buffer is accessible through the NIC-Host interconnect, where
the NIC inbound engine is connected. Only the NIC inbound engine can write through this port, hence it gets the full write bandwidth. Other units connected to the NIC-Host interconnect that can
access the L2 packet buffer, namely the NIC outbound engine and the off-cluster DMA engine, share the read bandwidth. 
The second port is connected to both DMA and PE interconnects. 
This configuration allows supporting a maximum line rate of 512 Gbit/s, making
\name suitable for up to 400 Gbit/s networks.

\mypar{L2 handler and program memory}
The L2 handler memory is less bandwidth-critical than the L2 packet buffer, but not less important. In the current configuration, the handler memory is 4 MiB. The sPIN programming model allows the host to access memory regions on the NIC to, e.g., write data needed by the handlers or read data back when a message is fully processed. Host applications can allocate memory regions
in this memory through the NIC driver, which manages the allocation state. The host can copy data in the handler memory before packets triggering handlers using it start arriving. For example, Di Girolamo et al.~\cite{spin-ddt} use this memory to store
information about MPI datatypes, deploying general handlers that process the packets
according to the memory layout described in the handler memory. 
Differently from the packet buffer, we foresee that the handler memory can be targeted more frequently by the HPUs with 32-bit word accesses, hence 
we adopt 64 bit-wide banks to reduce the probability of bank conflicts. 
Similarly, to the L2 packet buffer, the handler memory can be involved by flows 2 and 3 and offers a maximum bandwidth of 512 Gbit/s per port, full duplex.

The program memory (32 KiB) stores handlers code. It is accessed by the host to offload code and by the PE interconnect to refill the per-cluster 4 KiB instruction cache. Since this memory is not on the critical path, we implement it as single-port, half-duplex, with 64 Gbit/s bandwidth. 
\review{The per-cluster instruction cache is 4-way set associative with 8 ports. The concept of the home cluster, which tries to schedule packets of the same message (i.e., same handlers' code) to the same cluster, helps to reduce instruction cache pollution.}{E.1}

%The NIC inbound engine has access to the L2 packet buffer through
%the NIC-Host (NHI) interconnect. All interconnects in \name are based
%on the AXI4 protocol~\salvo{cite}. The NHI interconnect connects the NIC
%inbound engine and the L2 packet buffer with 512 bit (data-width) ports. This,
%o

%\clearpage
%\vspace{-0.5em}
\subsection{NIC integration}\label{sec:integration}
\vspace{-0.3em}

%\salvo{todo: use execution context}
We described \name within the context of the NIC model discussed in
Section~\ref{sec:architecture} but, \emph{how to integrate a \name unit in
existing networks?} To answer this question, we identify a set of NIC
\emph{capabilities}, some of which are required for integrating \name, and
others that are optional but can provide a richer handler semantic. The
required capabilities are:
%
%\begin{itemize}[noitemsep,topsep=0pt,parsep=0pt,partopsep=0pt,leftmargin=*]

\mylist{Message/flow matching} Packet handlers are defined per message/flow on the receiver side. The NIC must match a packet to a message/flow to identify the 
handler(s) to execute. We do not explicitly define messages or flows because this depends on the network where \name is integrated into. For \name, a message or flow is a sequence of packets targeting the same message processing queue (MPQ, see Section~\ref{sec:control}). The feedback channel to the NIC inbound engine is used to communicate when an MPQ becomes idle 
%(i.e., does not contain enqueued HERs) 
and can be remapped to a new NIC-defined message or flow.

\mylist{Header first} The first packet that is processed by \name
must carry the information characterizing the message. 
This requirement can be relaxed if packets carry information to identify a message or flow (e.g., TCP, UDP).

%\enlargethispage{0.6\baselineskip}
%If a network
%defines a message as a sequence of packets where only the first packet contains
%a message header, then the first packet must be delivered first. 
%\end{itemize}
%
NICs can provide additional capabilities that can (1) extend the functionalities
that the handlers have access to and (2) let the applications make stronger
assumptions on the network behavior. Applications can query the NIC
capabilities, potentially providing different handlers depending on the
available capabilities.  One such capability is \emph{reliability}.
With a reliable network layer,
\name is guaranteed to receive all packets of a message and to not receive
duplicated packets. With this capability, applications
can employ non-idempotent handlers. Otherwise, the handlers have to take
into account that, e.g., they can be executed more than once on the same packet.

\subsubsection{Match-action tables}
%\vspace{-0.5em}

NICs providing match-action table abstraction~\cite{pontarelli2019flowblaze, flexnic, bosshart2014p4} are an ideal candidate for a \name integration. With this abstraction, users can install packet parsing rules that lead to specific actions. To integrate \name, a new action should be made available that has the effect of forwarding the matched packet to the \name unit, together with an execution context that is associated with the match-action entry. 
\review{
This solution enables applications to define their own concept of message or flow, providing the greatest flexibility. For example,
applications can define a flow as a TCP stream (i.e., by matching on both IP and TCP headers) or as all UDP packets targeting a specific port. This solution would not be affected by ossification because the way flows are defined can be programmed. 
For example, applications using HTTP/2~\mbox{\cite{belshe2015hypertext}} that multiplex multiple streams within the same long-lived TCP connection can define a \name{} message/flow as a single stream (i.e., matching on the HTTP/2 header).
Similarly, transport protocols like QUIC~\mbox{\cite{iyengar2018quic}} can match \name{} messages/flows on single streams of long-lived connections. 
}{C.2}
%\name can be integrated into NICs proving match-action tables~\cite{pontarelli2019flowblaze, flexnic, bosshart2014p4}.
%These tables allow the user to specify a set of rules to which
%packets can be matched. If a packet matches a certain rule, a specific action can be executed on this packet. Integrating \name in this context would mean to introduce a new action consisting of forwarding the packet to the \name unit with a given execution context. This approach would offer the greatest flexibility for packet matching by not being tied to a specific network protocol and allowing applications to arbitrarily define their concept of \emph{message} or \emph{flow}. 

%\vspace{-0.5em}
\subsubsection{RDMA-Capable Networks}

%\enlargethispage{0.5\baselineskip}

Remote Direct Memory Access (RDMA) networks let applications expose memory
regions over the network, enabling remote processes to access them for reading or writing data. 
When using RDMA, applications register memory regions on the NIC, so that its IOMMU can translate virtual to physical addresses. Whenever a
remote process wants to, e.g., perform a write operation, it has to specify where in the target memory the data has to be written. This memory location can be directly specified by its target virtual memory address in the write 
request~\cite{infiniband2000infiniband, alverson2012cray}, or
indirectly~\cite{portals42}. In the indirect case, the application not only
registers the memory but also specifies a receive descriptor that can be
matched by incoming remote memory access requests: e.g., in Portals 4, these
descriptors are named list entries or matched list entries according to
whether they are associated with a set of matching bits or not. 

In general, RDMA NICs already perform the packet matching on the NIC.
In the direct case, the NIC matches the virtual address carried by the
request to a physical address. In the indirect case, the NIC matches the
packet to the receive descriptor, to derive the target memory location.
Hence, the required \emph{message matching} capability is provided;
the question is: to which object do we attach the \name handlers?
Table~\ref{table:rdma-nets} reports different RDMA-capable networks and objects where the \name handlers can be attached. For example,
associating handlers to the InfiniBand queue pair means that all packets targeting that queue pair will be processed by \name.

\begin{table}[t]
\scriptsize
\centering
\setlength{\tabcolsep}{4pt}
\begin{tabular}{llc}
\textbf{Network} & \textbf{Interface} & \textbf{Handlers Descriptor} \\
\toprule
InfiniBand~\cite{infiniband2000infiniband}, RoCE~\cite{subramanian2012remote} & ibverbs~\cite{bedeir2010building} & Queue Pair \\
Bull BXI~\cite{derradji2015bxi}, Cray Slingshot~\cite{sensi2020indepth} & Portals 4\cite{portals42} & Match List Entry \\
Cray Gemini~\cite{alverson2010gemini}, Cray Aries~\cite{alverson2012cray} & uGNI, DMAPP~\cite{ugnidmapp} & Memory Handle \\
\bottomrule
\end{tabular}
\vspace{0.4em}
\caption{RDMA networks and sPIN handlers attach points.}
\label{table:rdma-nets}
\vspace{-3.5em}
\end{table}
%QP -> Infiniband, RoCE. (ibverbs)
%LE/ME -> Bull BXI, Slingshot (Portals 4 like)
%Memory Handle -> Cray Gemini, Cray Aries (uGNI, DMAPP)

The second required capability is \emph{header first}. For
InfiniBand, this is given by the in-order delivery that the network already
provides. For other networks that cannot guarantee that, the NIC must be able to buffer or
discard payloads packets arriving before the header packet.
RDMA-capable networks already implement reliability at the network layer, hence
applications can adopt non-idempotent handlers. 

\subsection{NIC driver}
\vspace{-0.3em}

To expose packet-processing functionalities, the NIC driver
needs to implement the sPIN interface described by Hoefler et al.~\cite{spin}. 
In particular, the driver manages the NIC memory by letting applications allocate
memory regions for data (e.g., handler memory) and code (e.g., program memory). 
The \name unit is not involved in applications memory management, which is
delegated to the software layer. A detailed description of a NIC driver is out of the scope of this work.

%\vspace{-0.5em}
\subsection{Special cases and exceptions}\label{sec:discussion}
\vspace{-0.3em}

%\enlargethispage{0.5\baselineskip}

%\noindent\emph{What happens if \name cannot sustain line rate?} Strategies to
%tackle this situation are already described in previous works~\cite{spin,
%spin-ddt}.  In \name, the control processor detect the bottlenecks by
%looking at the frequency of completion feedbacks. The exact actions to take
%mainly depend on the network where \name is integrated into: e.g., killing a
%connection~\cite{portals42}, applying back pressure to slow down the
%senders~\cite{infiniband2000infiniband}, or sending explicit congestion
%notification messages~\cite{ramakrishnan2001addition}. 
%
%To avoid this situation, the host could enforce limitations on the offloaded handlers
%(similaraly to eBPF~\mbox{\cite{miano2018creating}}): e.g., bounded loops, limited instructions
%count, static handlers analysis. 

\noindent\emph{Can \name deadlock if no processing cluster can accept new tasks?}
In this case, the task dispatcher will block, waiting for a queue to become available again and this will create back-pressure towards the NIC inbound engine.
The system cannot deadlock because the processing clusters can keep
running since they are not dependent on new HERs to arrive. The header-before-payloads
dependency does not cause problems because if payload handlers are waiting
for the header, then it is guaranteed that the header is being already processed
(because of the \emph{header-first} requisite and the in-order scheduling guaranteed by the MPQ engine on a per-message basis). 
If badly-written handlers deadlock, the HPU driver watchdog will trigger causing the handler termination.

\vspace{0.3em}
\noindent\emph{What if a message is not fully delivered?} The
completion feedback will not be triggered causing resources (e.g., message state in the MPQ engine) to not be freed. \name can
detect this case and force resource release (see Section~\ref{sec:control}). 

\vspace{0.3em}
\noindent{\emph{How is encrypted traffic handled?} 
Handlers are responsible for the decryption of incoming data. We foresee the possibility
of supporting user handlers with libraries providing common functions like crypto primitives. 
Given the modular design of \name, a per-cluster crypto engine can be deployed to enable
hardware-accelerated crypto primitives (e.g., AES-EBC). While a crypto engine for PULP (hence \name-compatible) already exist~{\cite{gurkaynak2017multi}}, we consider its evaluation as future work.}

%\begin{itemize}
%\item Discuss how we handle the cases where we detect being the bottleneck. 
%\item HW vs SW scheduling
%\item discuss possible deadlocks and how to fix them. Just dropping packets
%may not work because then the NIC satisfied header-first, but we drop the header. 
%So it would be legal for the NIC to forward payload packets. We need something stronger:
%we drop the header and kill the connection: now the NIC knows that is has to drop 
%all the packets for that connection and also signal the sender that the connection is invalid. 
%\item Constraint the resources. For example, given the number of SMDs, how many PMDs and other resources do we need?
%\item solutions like inloading when we are the bottlneck work well with sockets (we can
%flag a packet as processed or not), but don't work with RDMA: what if our handlers start
%writing f(pkts) to main memory, then we inload, and the NIC will just append the rest
%of the pkts to the same memory? Do we need something like a shadow buffer? 
%\end{itemize}

%\clearpage
%\vspace{-0.5em}

\section{Evaluation}\label{sec:evaluation}
\vspace{-0.4em}

Our evaluation aims to answer the following questions: (1) How big is \name in terms of post-synthesis area and how does that scale with the number of HPUs? (2) In which cases can \name sustain line rate? (3) Does the choice of implementing sPIN on top of a RISC-V based architecture with a flat non-coherent memory hierarchy pay off? What are the trade-offs of choosing more complex architectures for sPIN?

\mypar{Simulation environment} We simulate PsPIN in a \textbf{cycle-accurate testbench comprised of SystemVerilog modules}. \textbf{We use synthesizable modules for all \name components. We develop simulation-only modules modeling the NIC inbound and outbound engines.} Our inbound engine takes a trace of packets as input and injects them in \name at a given rate. The outbound engine reads data from \name according to the received commands, generating memory pressure. The host interface is emulated with a PCIe model (PCIe~5.0, 16 lanes), implemented as a fixed-rate data sink. Unless otherwise specified, we do not limit the packet generator injection rate in order to test the
maximum throughput \name can offer. Packet handlers are compiled with the PULP SDK, which contains an extended version of GCC 7.1.1 (riscv32). All handlers are compiled with full optimizations on (\texttt{-O3 -flto}).

%\vspace{-0.5em}
\subsection{Hardware Synthesis and Power}\label{sec:hw-synthesis}
\vspace{-0.3em}

We synthesized \name in GlobalFoundries' 22\,nm fully depleted silicon
on insulator (FDSOI) technology using Synopsys DesignCompiler 2019.12, and we were able
to close the timing of the system at 1\,GHz. 
\review{We employ Invecas' memory compiler to generate SRAM macros that are tailored to the architectural requirements.}{E.2}
Area and power measurements are summarized in Table~\ref{tbl:area_power}.
Including memories, the entire
accelerator has a complexity on the order of 95\,MGE.\footnote{One gate
equivalent (GE) equals 0.199\,$\mu m^2$ in
GF 22\,nm FDSOI.} Of the overall area, the four clusters (including their L1 memory and the intra-cluster scheduler) occupy 43\,\%, the L2 memory 51\,\%, the inter-cluster scheduler 3\,\%, and the inter-cluster interconnect and L2 memory controllers another 3\,\%.  The L2
memory macros occupy a total area of 9.48\,mm$^2$.  
Depending on the NIC architecture where \name is integrated into, the L2 packet buffer could be mapped to the NIC packet buffer, saving memory area.
The area of the clusters is
dominated by the L1 memory macros, which take 1.65\,mm$^2$ per cluster.  The
instruction cache and the cluster interconnect have a complexity of ca.\
700~kGE per cluster, which corresponds to ca.\ 0.2\,mm$^2$ at 70\,\% placement
density.  Each core has a complexity of ca.\ 50~kGE, which corresponds to ca.\
0.014\,mm$^2$.  The total cluster area is ca.\ 1.99\,mm$^2$.  The total area of our
architecture is ca.\ 18.5\,mm$^2$ (\textbf{S6}). For comparison, from~\cite{mair20164, pyo201523} it can be inferred that a Mellanox BlueField SoC, scaled to 16 ARM A72 cores (22 nm), would occupy ca. $51$\,mm$^2$.

\begin{table}[h]
	\vspace{-0.5em}
	\scriptsize
	\begin{tabular}{lrrrrrr}
		\multirow{2}{2cm}{\textbf{Component}} & \multicolumn{3}{c}{\textbf{Area} (mm$^2$)} & \multicolumn{3}{c}{\textbf{Power} (W)} \\
		& {Unit} & {Total} & {Perc.} & {Unit} & {Total} & {Perc.} \\
		\toprule
		\textbf{\name{}} & 18.47 & 18.47 & 100.0\% & 6.08 & 6.08 & 100.0\% \\
		$\drsh$ \textbf{L2 memories} ($\times$1) & 9.48 & 9.48 & 51.3\% & 1.09 & 1.10 & 18.1\% \\
		$\drsh$ \textbf{Interconnect} ($\times$1) & 0.57 & 0.57 & 3.0\% & 0.71 & 0.71 & 11.7\% \\
		$\drsh$ \textbf{Cluster} ($\times$4) & 1.99 & 7.95 & 43.0\% & 0.94 & 3.77 & 62.0\% \\
		~$\drsh$ \textbf{L1} ($\times$1) & 1.65 & 1.65 & 82.9\% & 0.52 & 0.52 & 55.3\% \\
		~$\drsh$ \textbf{Core} ($\times$8) & 0.01 & 0.08 & 4.0\% & 0.02 & 0.14 & 15.3\% \\
		~$\drsh$ \textbf{Instr. cache} ($\times$1) & 0.08 & 0.08 & 4.0\% & 0.14 & 0.14 & 15.1\% \\
		~$\drsh$ \textbf{Interconnect} ($\times$1) & 0.06 & 0.06 & 3.0\% & 0.11 & 0.11 & 11.3\% \\
		\bottomrule
	\end{tabular}
	\vspace{0.5em}
	\caption{Area and energy of main \name{} components. Percentages are relative to the parent component.}
	\label{tbl:area_power}
	\vspace{-1.8em}
\end{table}

We derive a worst-case upper bound for the power consumption of our architecture by
assuming 100\,\% toggle rate on all logic cells and 50/50\,\% read/write
activity at each memory macro.  The overall power envelope
is 6.1\,W, 99.8\,\% of which is dynamic power (\textbf{S6}).  The four clusters consume 62\,\% of the total
power, ca.\ 3.8\,W.  Within each cluster, the L1 memory consumes ca.\ 55\,\% of
the power.  The L2 memory consumes 18\,\% of the total power, ca.\ 1.1\,W.
The inter-cluster scheduler consumes 8\,\% of the total power, ca.\ 0.5\,W.
The inter-cluster interconnect and L2 memory controllers consume 11.7\,\%, ca.\
0.7\,W.  As our architecture offers 32~HPUs, the power normalized to the number
of HPUs is 190\,mW.
%The actual power consumption will be significantly smaller for most practical applications, but measuring it will only be possible once a physical prototype can be tested extensively.
%\salvo{can we add a plot showing an estimate of how the area scales with number of HPUs/clusters? Should we also vary the number of HPUs per cluster?  Andreas: Is this still relevant, do we want to do it for this deadline?  We already reach 400 Gbit/s, so which other performance targets do we have?  If yes, we could derive numbers with modest scaling factors from the area spread sheet.  Torsten: not important}

%\vspace{-0.5em}
%\enlargethispage{\baselineskip}
\subsection{Microbenchmarks}
\vspace{-0.3em}

We now investigate the performance characteristics of \name: we first discuss the latencies
experienced by a packet when being processed by \name. Then, we study the maximum packet
processing throughput that \name can achieve and how the complexity of the packet handlers can affect it. 

%\vspace{-0.5em}
\subsubsection{Packet Latency}\label{sec:latency}
%\vspace{-0.5em}

We define the packet latency as the time that elapses from when \name receives an HER from the NIC inbound
engine to when the completion notification for that packet is sent back to it. It does not include the time needed by the NIC inbound engine to write the packet to the L2 packet buffer. The measurements
of this section are taken in an unloaded system by instrumenting the cycle-accurate simulation. Overall, we observe
latencies ranging from 26 ns for 64 B packets to 40 ns for 1024 B ones. In particular, a task execution
request takes 3 ns to arrive to the cluster-local scheduler (i.e., CSCHED in Figure~\ref{fig:cluster-detail}). At that point, the packet is copied to the cluster L1 by the cluster-local DMA engine. This transfer has latencies varying from 12 ns for 64 B packets to 26 ns for 1024 B packets. Once the data reaches L1, the task is assigned to an HPU driver in a single cycle. The HPU runtime takes 7 ns to invoke the handler: this time is used for reading the handler function pointer, setting up the handler's arguments, and making the jump. Once the handler completes, the runtime makes a single-cycle store to the HPU driver to inform it of the completion. The completion notification takes 1 ns to get back to the NIC inbound engine, but it can be delayed of additional 6 ns and 2 ns in case of the round-robin arbiters prioritize other HPUs and clusters, respectively.

%\vspace{-0.5em}
\subsubsection{Packet processing throughput} 
%\vspace{-0.5em}

In Section~\ref{sec:datapath} we describe three critical data flows that can run over a \name unit. Flow 1 (\emph{inbound flow}) moves data from the NIC inbound engine to the L2 packet memory and, from there, to the L1 memory of the processing cluster where the packet has been assigned.
Moving packet data to L1 memories is not always needed. For example, a handler might only use the packet header (e.g., filtering), the packet header plus a small part of the packet payload (e.g., handlers looking at application-specific headers), or they might not need packet data at all (e.g., packet counting). 
Applications specify the number of bytes that handlers need for each packet.
Flows 2 and 3 move data from \name to the outbound interfaces, namely the
NIC outbound (\emph{outbound NIC flow}) and the host interface through PCIe (\emph{outbound host flow}). They are generated by the handlers, which can issue commands to move data to the NIC or to
the host. Handlers do not necessarily issue commands as they can directly consume data
and communicate results to the host once the message processing finishes: e.g., handlers performing data reductions on the NIC, letting the completion handler write data to the host.

\begin{figure}[t]
	\vspace{-0em}
	\centering
	\includegraphics[width=\columnwidth, trim=0 0 5 0]{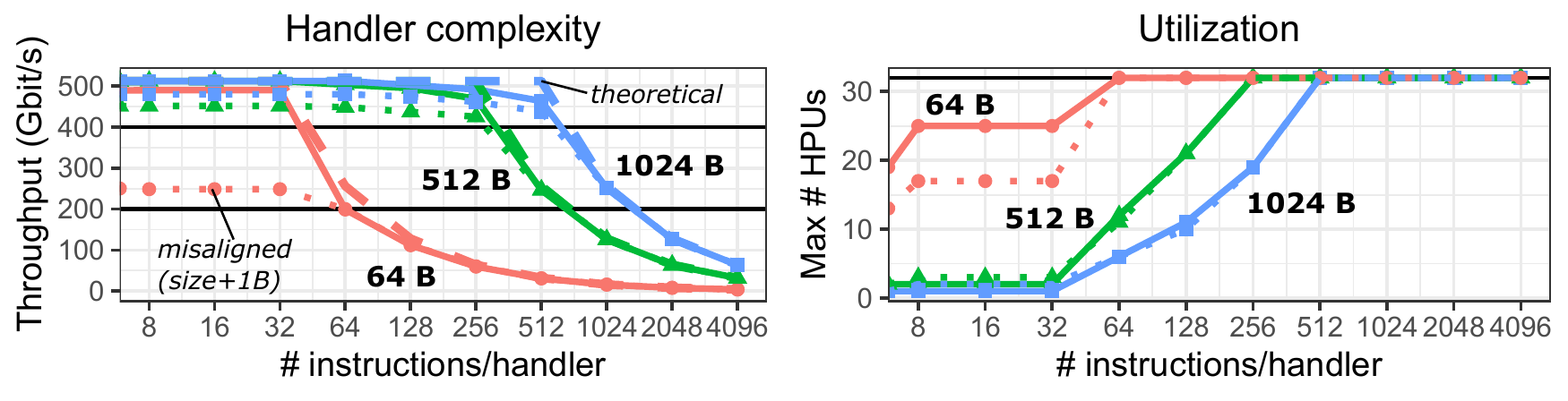}
	\vspace{-2.4em}
	\caption{\name maximum throughput.}
	\label{fig:hcomplexity}
	\vspace{-2em}
\end{figure}

\mypar{Inbound flow} We measure the throughput \name can sustain for the inbound flow. We measure
it as function of the frequency of the completion notifications received by the MPQ engine and
the packet size. 
Figure~\ref{fig:hcomplexity} (left) shows the throughput for
handlers executing different number of instructions
(x-axis) and for different packet sizes (i.e., 64 B, 512 B, and 1024 B packets). We also
include: (1) the maximum throughput that the \name can achieve: this is the minimum between
the interconnect bandwidth and the cumulative bandwidth offered by the 32 HPUs when
executing $x$ instructions; and (2) the throughput for misaligned packets (i.e., \mbox{packet size + 1 byte}). We let each handler execute $x$ integer arithmetic
instructions, each completed in a single cycle. The x-axis can also be read as handler duration
in nanoseconds. The data shows how \name can schedule aligned packets at the maximum available
bandwidth and the HPU runtime introduces minimum overhead (i.e., 8 cycles per packet, see Section~\ref{sec:latency}).

Figure~\ref{fig:hcomplexity} (right) shows the maximum number of HPUs that are utilized when running
handlers executing $x$ instructions, for different packet sizes. \name can schedule one 64 B packet per cycle. Even with empty handlers, we need 19 HPUs to process them because of the overhead necessary to invoke the handlers. With bigger packets, the time budget increases: handlers with small instruction counts can process 512 B and 1024 B packets at full throughput with a single HPU.

\begin{figure}[H]
\vspace{-1.2em}
\centering
\includegraphics[width=\columnwidth, trim=0 0 5 0]{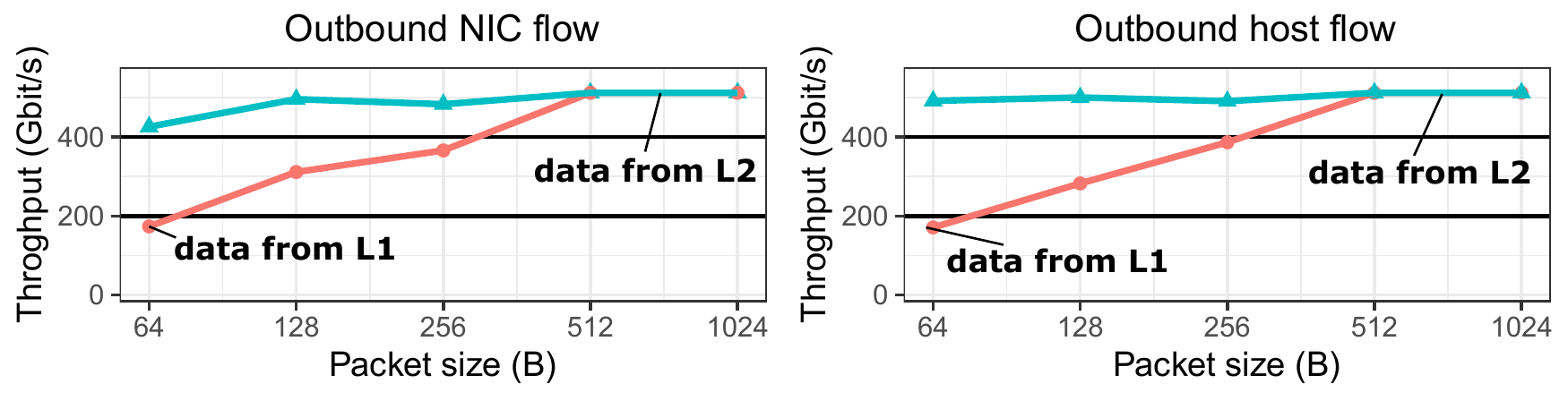}
\vspace{-2.4em}
\caption{Moving data out of \name. }
\label{fig:outbound-throughput}
\vspace{-1em}
\end{figure}

\mypar{Inbound + outbound flows}
We now study the throughput offered when packets are received and sent out of \name. The execution context is configured to move the full packet to L1. For testing the outbound NIC flow, we develop handlers
performing a UDP ping-pong: they swap source and destination IPs and UDP ports, then issue a NIC command to send it back over the network. Overall, this handler consists of 
27 instructions (20 for the swap and 7 for the issuing the command). The handlers for the outbound host flow only issue a DMA command to move the packet to the host, without modifying it. 
Figure~\ref{fig:outbound-throughput} shows both the cases in which the packet is sent from L1 and from the L2 packet buffer. 

%Figure~\ref{fig:outbound-throughput} shows the results of this benchmark. 
The L2 packet buffer, with its 32 512-bit-wide banks, is optimized for wide accesses, as the ones performed by the DMA engines of the involved units. The L1 TCDM is optimized for serving 32-bit word accesses from the HPUs and organized in 64 32-bit-wide banks. 
This difference shows up in the throughput and it is caused by a higher number of bank conflicts in the data-from-L1 case: with 64 B packets, both the outbound flows hardly reach 200 Gbit/s when reading from L1, while 400 Gbit/s is reached when reading data from the L2.
For bigger packets ($\ge$512 B), the time budget is large enough to allow also the L1 case to reach full bandwidth.

\begin{figure}[t]
	\centering
	\includegraphics[width=\columnwidth, trim=0 0 0 0]{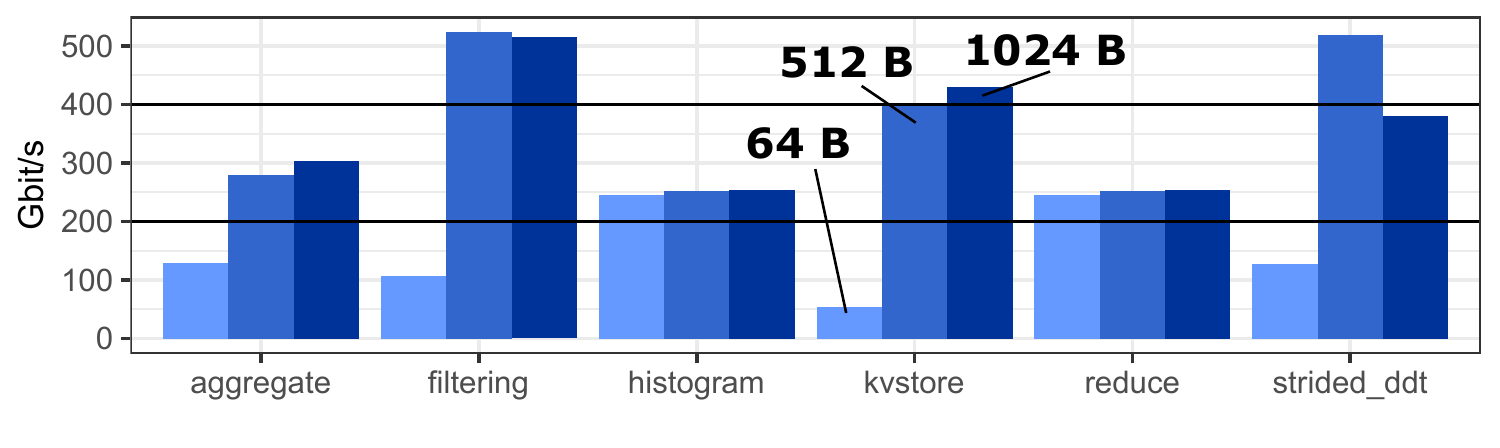}
	\vspace{-2.8em}
	\caption{Handler throughput on \name for different packet sizes.}
	\label{fig:apps-pspin}
	\vspace{-1.7em}
\end{figure}

%\vspace{-0.5em}
\subsection{Handlers Characterization}\label{sec:apps}
\vspace{-0.2em}

To evaluate the performance of \name for realistic packet handlers, we select 
a set of use cases ranging from packet steering to full message processing.
We first show the throughput they can
achieve on \name, then we measure the per-core throughput achievable on \name{} {RISC-V}
and compare it to the one achieved on more complex
and powerful, but bigger, architectures such as x86 and ARM.
\review{This comparison aims to analyze costs and benefits of employing more complex architectures for packet processing and to motivate our design choice of employing simple RISC-V cores as HPUs.}{D.2}
We simulate a network with zero inter-packet-delay, in order to not make
our results network-bound and show the maximum achievable throughput. 
The considered use cases are:

\mylist{Data reduction} Reducing data of multiple messages is a
core operation of collective reductions~\cite{mpi-3.0} and one-sided
accumulations~\cite{hoefler2015remote}. Given $n$
messages, each carrying $m$ data items of type $t$, this operation computes an
array of $m$ entries of type $t$  where entry $i$ is the reduction
of the $i$-th data item across the $n$ messages.
We benchmark an instance of this use case (named \emph{reduce}) with 512
packet, each carrying 512 32-bit integers. Payload handlers accumulate
data in L1 using the sum operator. 
The completion handler informs the host that the result is available with a
a direct host write command. 
%Alternatively, the result can be directly DMAed to host memory.

%\salvo{We get multiple messages. Each message is
%composed by multiple data items (can spread on multiple packets). Now we want
%to accumulate per data item: acc[data\_idx] = msg\_1[data\_idx] + ... +
%msg\_n[data\_idx]. Vary data items. Maybe it does not make sense to show again
%what happens when using L2 or L1 as accumulator (discuss). Check if we can use
%RISC-V SIMD.}

\mylist{Data aggregation} 
Utilized in, e.g., data-mining applications~\cite{kumar2012horizontal}, 
this operation consists in accumulating the
data items carried by a message. This benchmark
(\emph{aggregate}) uses a 1 MiB message of 32-bit integers that are summed up
in L1. The completion handler copies the
aggregate to host memory.

%\salvo{We get a large message and we want to
%sum up all the data items that it carries. Data items: int32, in16, int8.
%Variants: accumulate in L1 and in L2. Options: always carry the same packet,
%maybe show what happens if you leave it in L2.}

\mylist{Packet filtering/rewriting} Typical of intrusion-detection, 
traffic monitoring, and packet sniffing systems~\cite{deri2007high}. For each packet, it queries
an application-defined hash table (in L2) by using the source IP
address (32-bit) as key. If a match is found, the UDP destination port is overwritten with
the matched value and
written to host memory. This benchmark, named \emph{filtering}, uses 512 messages and
a hash table of 65,536 entries. 

%\salvo{Keep a hash table of flows in L2. We
%reprent a flow by (source rank, tag). For each message we get, we want to query
%the hash table and, if the flow is present, make a single write to L2 to
%rewrite the tag. Use BSW instructions. How to build the hash table? Can we
%cache parts of the hash table to L1?}

\mylist{Key-Value cache} A key-value store (\emph{kvstore})
cache on the NIC. The cache is stored in L2 and is implemented as a
set-associative cache to limit the L2 accesses needed to maintain 
the cache (e.g., eviction victims are chosen within a row). We generate a
YCSB~\cite{ycsb} workload of 1,000 requests (50/50 read/write ratio, $\theta$=1.1).
The cache associativity is set to 4 and the total number of entries is set to
500. The set is determined as the key (32-bit integer) modulo the number of sets. 

%\salvo{Keep hash table in L2. Generate workload
%of write/reads to get as packets. Use the hash table as key/value store.}.

%\subsubsection{Data Decryption} \salvo{Discuss with Konst.}

\mylist{Scatter} This use case (\emph{strided\_ddt}) models data transfers that
are copied to the destination memory according to a receiver-specified
memory layout~\cite{mpi-3.0, spin-ddt}. This benchmark sends a 1 MiB message 
that is copied to host memory in blocks of 256 bytes and with a stride of 512 bytes.  
The layout description (i.e., block size and stride) is stored in L2.

%\enlargethispage{\baselineskip}

\mylist{Histogram} Given a set of messages, it summarizes 
data items by value. This application is common in
distributed join algorithms~\cite{barthels2017distributed}. In our instance, we
receive 512 messages, each carrying 512 integers randomly generated in the $[0,
1024]$ interval. The handlers count how many data items per value have been
received and finally copy the histogram to the host.

%\enlargethispage{\baselineskip}
%\vspace{-0.5em}
\subsubsection{Handler Throughput}
%\vspace{-0.5em}

%To compute the area efficiency, we consider the single handler execution time
%reported in Figure~\ref{fig:apps} and divide it by the core area of the
%specific architecture.  

Figure~\mbox{\ref{fig:apps-pspin}} shows the throughput achieved by the
considered handlers on \mbox{\name} for different packet sizes. 
We observe that \name achieves 400 Gbit/s for \emph{filtering}, \emph{kvstore}, and \emph{strided\_ddt} already for 512 B packets. In the other cases, handlers are compute-intensive, and they operate on every 32-bit word of each received packet. Nonetheless, \name achieves more than 200 Gbit/s, which the state-of-the-art network speed, from 512 B packets. Thanks to the modularity of this architecture (\textbf{S8}), a scenario where 400 Gbit/s must be sustained also for this type of workload can be satisfied by doubling the number of processing clusters.

%Differently from Figure~\mbox{\ref{fig:apps-max-throughput}}, which shows the throughput computed from the handler execution times, in this case the full system is utilized and the throughput is measured as function of the completion notifications. In these settings, handlers experience scheduling overheads, contention on memories, NIC outbound engine and off-cluster DMA engine. 

\begin{figure}[t]
	\centering
	\includegraphics[width=\columnwidth, trim=0 0 0 0]{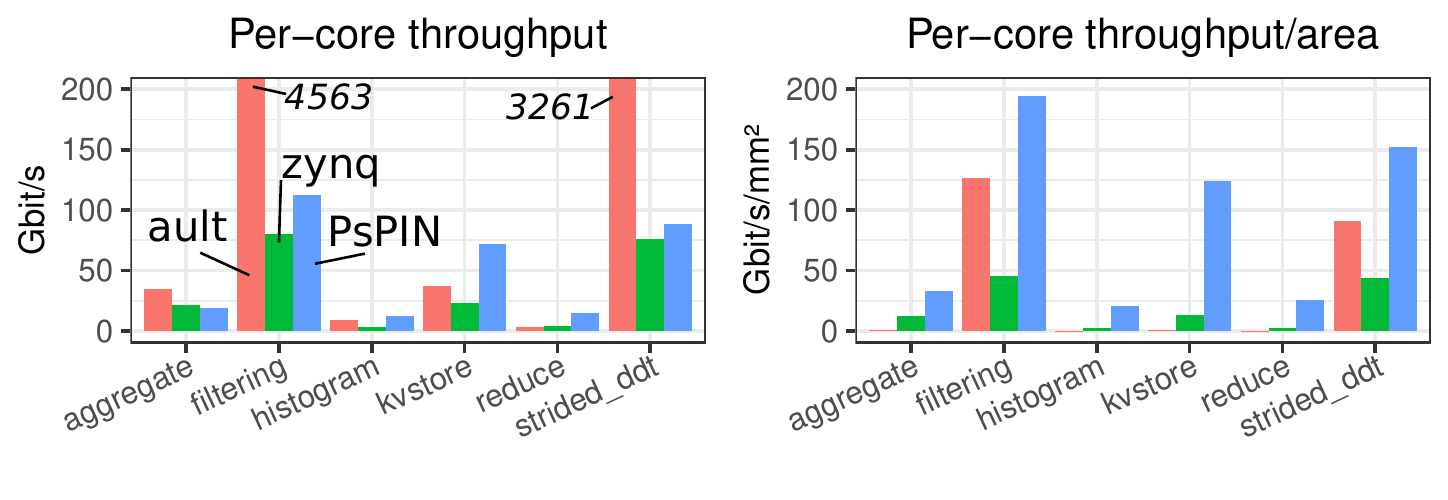}
	\vspace{-3.2em}
	\caption{\review{Per-core handler throughput: ARM vs x86 vs RISC-V.}{E.4}}
	\label{fig:apps-max-throughput}
	\vspace{-1.8em}
\end{figure}

%\begin{table}[h]
%	\scriptsize
%	\vspace{-1em}
%	\centering
%	\setlength{\tabcolsep}{4.5pt}
%	\begin{tabular}{lcccccc}
%		\textbf{Arch.} & \textbf{Tech.} & \textbf{Die area} & \textbf{Cores} & \textbf{Memory} & \textbf{Area/core} & \textbf{Area/core (scaled)}\\
%		\toprule
%		ault & 14 nm & 485 mm$^2$~\cite{ault-area} & 18 & 43.3 MiB & 17.978 $mm^2$ & 35.956 mm $^2$\\
%		zynq & 16 nm & 3.27 mm$^2$~\cite{arm-area} & 4 & 1.125 MiB & 0.818 mm$^2$ & 1.752 mm$^2$ \\
%		\textbf{\name} & 22 nm & 19.8 mm$^2$ & 32 & 12 MiB & 0.619 mm$^2$ & 0.619 mm$^2$\\
%		\bottomrule
%	\end{tabular}
%	\caption{Architectural characteristics.}
%	\label{tbl:ap-estimates}
%\end{table}

%\vspace{-0.5em}
\subsubsection{RISC-V vs x86 vs ARM}
%\vspace{-0.5em}

This set of experiments outlines the benefits of adopting a simple, 
RISC-V-based architecture over more powerful and complex ones, such as
x86 and ARM. 
We select two representative architectures, showing not only the effects
of different CPU types but also of different memory subsystem configurations (e.g., caches vs scratchpads):
\begin{itemize}[noitemsep,topsep=0pt,parsep=0pt,partopsep=0pt,leftmargin=*]
	
	\item \textbf{ault} 18-core 64-bit 2-way SMT, 4-way superscalar, out-of-order-execution Intel Skylake Xeon Gold 6154 (3\,GHz). 
	
	\item \textbf{zynq} Xilinx Zynq ZU9EG MPSoC featuring a quad-core ARM Cortex-A53 (64-bit 2-way superscalar at 1.2\,GHz).
	%The Cortex-A53 is a 64-bit 2-way superscalar processor (1.2\,GHz).
\end{itemize}
To run on these architectures, we develop a benchmark that loads a predefined
list of packets in memory, spawns a set of worker threads, and statically assigns
the packets to the workers. This setting can be compared to an ideal DPDK
execution since the packets are already in memory and the workers do not 
experience any DPDK-related overheads (e.g., polling device ports, copying bursts in local buffer).
If not otherwise specified, the packet size is set to 1 KiB.

\begin{figure*}[t]
	\centering
	\vspace{-1em}
	\includegraphics[width=\textwidth]{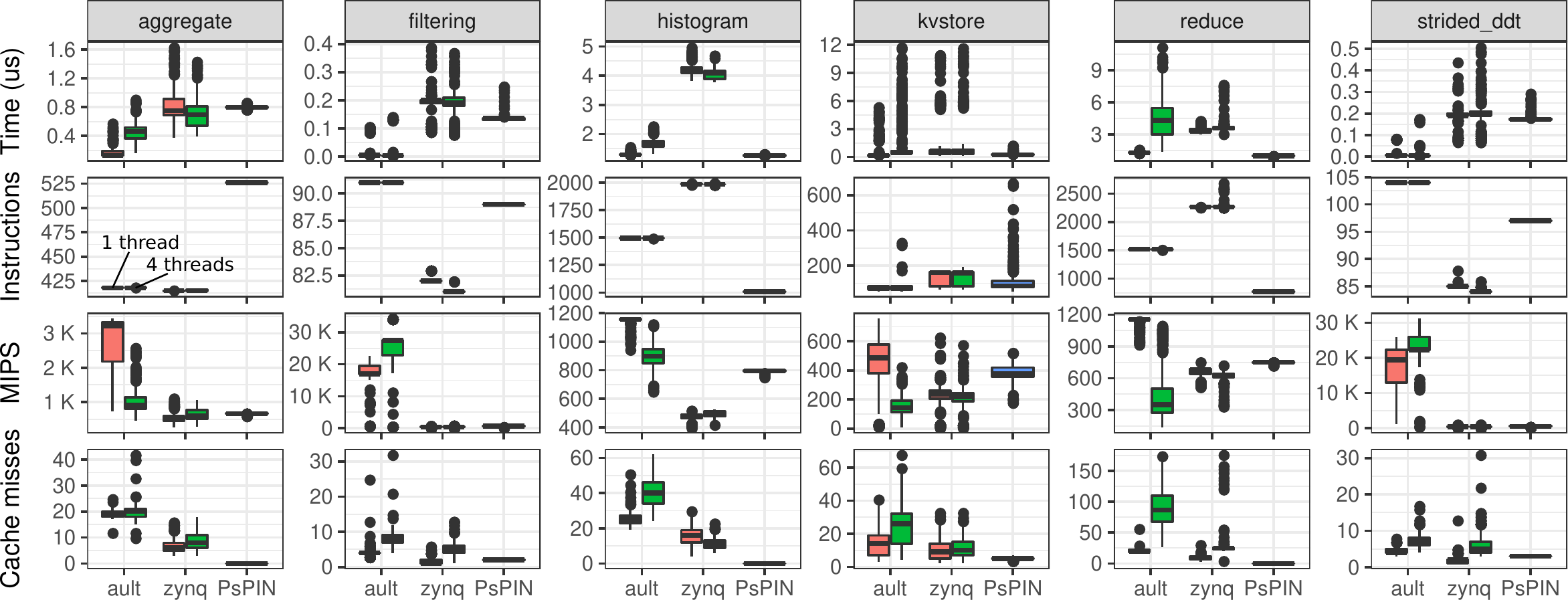}
	\vspace{-1.9em}
	\caption{
		\review{Handlers performance on different architectures. For zynq and ault, we show the cases with one and four worker threads. The upper and lower whiskers of the boxplot represent $Q3 + 1.5 \cdot IQR$ and $Q1 - 1.5 \cdot IQR$ ($Q$i: i-th quartile; $IQR$: inter-quartile range), respectively.}{E.4}}
	\label{fig:apps}
	\vspace{-1.7em}
\end{figure*}

\review{Figure~\mbox{\ref{fig:apps-max-throughput}} (left) compares the per-core throughput of each architecture, which is computed as function of the median handler completion time when using a number of worker threads equal to the number of available cores. 
Despite the fact that this comparison is disadvantageous for \name{} because it is the one potentially experiencing the most memory contention with its 32 cores, \name{} shows better per-core throughput over the best competitor for \emph{histogram} (1.3x), \emph{kvstore} (1.9x), and \emph{reduce} (5.3x). For less memory-bound workloads, the more powerful cores of ault and zynq outperform \name{} by up to 1.8x for \emph{aggregate}, 40x for \emph{filtering}, and 36x for \emph{strided\_ddt}.}{B.2, E.4}

\review{However, comparing the per-core throughput without factoring in the area of each architecture is not fair (e.g., ault is 26x larger than \name{}). Table~\mbox{\ref{tbl:ap-estimates}} summarizes area estimates and shows the scaled area per processing element (\emph{Area/PE (scaled)}), which is the area per processing element (\emph{Area/PE}) scaled to the same production process (22 nm) and same amount of memory per PE. 
Figure~\mbox{\ref{fig:apps-max-throughput}} (right) reports the throughput normalized to the scaled area for the considered architectures. \name{} is up to 10.7x more area-efficient than zynq (minimum: 2.6x for \emph{strided\_ddt}) and up to 248x more area-efficient than ault (minimum: 1.5x for \emph{filtering}) on all considered workloads.
We conclude that, while it is expected that more powerful architectures achieve higher raw throughput for compute-intensive workloads, \name{} provides better area efficiency and can sustain line rate while fully offloading packet processing to the NIC and freeing CPU resources.}

\newcolumntype{C}[1]{>{\centering\let\newline\\\arraybackslash\hspace{0pt}}m{#1}}
\begin{table}[h]
	\scriptsize
	\vspace{-1em}
	\centering
	\setlength{\tabcolsep}{2.9pt}
	\begin{tabular}{lcccccc}
		\textbf{Arch.} & \textbf{Tech.} & \textbf{Die area} & \textbf{PEs} & \textbf{Memory} & \textbf{Area/PE} & \textbf{Area/PE (scaled)}\\
		\toprule
		ault & 14 nm & 485 mm$^2$~\cite{ault-area} & 18 & 43.3 MiB & 17.978 mm$^2$ & 35.956 mm$^2$\\
		zynq & 16 nm & 3.27 mm$^2$~\cite{arm-area} & 4 & 1.125 MiB & 0.876 mm$^2$ & 1.752 mm$^2$ \\
		\textbf{\name} & 22 nm & 18.5 mm$^2$ & 32 & 12 MiB & 0.578 mm$^2$ & 0.578 mm$^2$\\
		\bottomrule
	\end{tabular}
	\vspace{0.3em}
	\caption{Architectural characteristics. PE: processing element}
	\label{tbl:ap-estimates}
	\vspace{-2.3em}
\end{table}

To gain more insights on the performance characteristics of the selected handlers, Figure~\ref{fig:apps} 
shows a set of performance metrics as measured on the considered architectures. 
We report handlers' execution times, number of executed instructions, MIPS (\emph{million instructions per second}), and cache
misses. For \name, L1 misses represent the number of accesses to either remote L1s or L2. For ault and zynq, performance is measured with CPU hardware counters~\cite{terpstra2010collecting}. 
To show the effects of resource contention, we run experiments with a single worker
thread (i.e., no contention) and with four workers in parallel.

%This benchmark shows that these architectures achieve similar handlers' execution times and
%architectural characteristics like hardware caches have a limited positive impact 
%for packet-processing workloads.

% 1) it's not that bad
\review{
For most of the considered use cases, running times on \name{} are not more than 2x the best case 
(i.e., no contention) of other architectures. 
The worst case is \emph{filtering}, which computes a hash function on an 8-byte value, resulting
in a compute-intensive task that allows ault to run this handler more than 30x times faster than \name{}.
In general, for workloads that mainly execute arithmetic instructions (e.g., \emph{aggregate} and \emph{filtering}) or do not frequently access shared or packet memory (e.g., \emph{strided\_ddt}), ault outperforms zynq and \name{} in terms of completion time. For example, on ault, the compiler optimizes
\emph{aggregate} by using SIMD packed integer instructions. However, as shown if Figure~\mbox{\ref{fig:apps-max-throughput}}, this difference does not take into account the larger area occupied by this architecture. 
Even though \name{} has a simpler architecture than ault and zynq, it still competes in overall execution time for the other cases due to the comparable rate of executed instructions per second (MIPS), which is influenced by higher L1 miss rates on ault and zynq (see \emph{histogram}, \emph{kvstore}, and \emph{reduce}). 
In \name{}, packets are copied directly into the L1 of the cluster where the handler is executed, enabling single-cycle access. Also, \name{} has no hardware caches, hence it does not suffer from cache-line ping-pong scenarios, as observed on other architectures for, e.g., \emph{histogram} and \emph{reduce}. \mbox{RISC-V} AMOs~\mbox{\cite{waterman2014risc}} enable single-cycle atomic operations that can save up to 3x instructions over other implementations (e.g., linked load, store conditional) for the \emph{reduce} and \emph{histogram} cases.
}{B.2}

%
% 3) in some cases we shine

\section{Discussion and future work}

The \name{} configuration and analysis that we show in this work is aimed at sustaining a 400 Gbit/s line rate.
\emph{How can PsPIN be scaled out to sustain higher bandwidths?} To reason about scaling we need to consider how to scale memories, interconnects, and cores and how this affects power and area. 
We identify two types of memories: the packet buffer, the size of which depends on the network bandwidth and the time packets spend in \name{} (i.e., the packet latency), and the handler memory, the size of which depends on the specific handlers that are offloaded (e.g., to store their state). Note that the second class does not depend on the network bandwidth.

%\enlargethispage{0.5\baselineskip}
\begin{figure}[h]
	\vspace{-1.2em}
	\centering
	\includegraphics[width=\columnwidth]{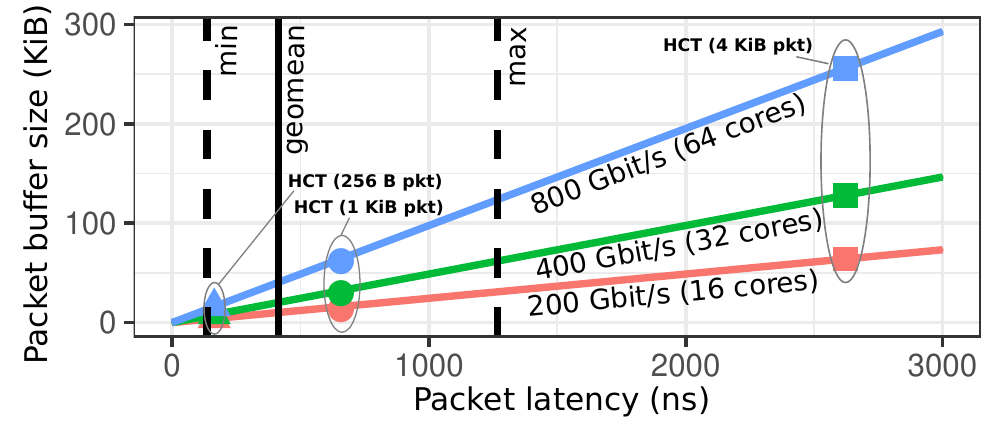}
	\vspace{-3em}
	\caption{Packet buffer size over packet latencies and line rates. Horizontal lines are min, geometric mean, and max handler running times among the ones of Figure~\ref{fig:apps}. Points indicate handler critical times (HCT) after which handlers are bottlenecks for a given packet-size/line-rate/number-of-cores combination.}
	\label{fig:scaling}
	\vspace{-0.5em}
\end{figure}

In Figure~\ref{fig:scaling} we use Little's law to determine the packet buffer size over different packet latencies (x-axis) and line rates (200 Gbit/s, 400 Gbit/s, and 800 Gbit/s). The packet latency is the time from when a packet arrives in \name{} to when it is processed and can be freed up from the packet buffer. As we show in Section~\ref{sec:latency}, the time needed to schedule a handler processing a given packet ranges from 26 ns for 64 B packets to 40 ns for 1024 B ones. For simplicity, we do not take into account the scheduling latency in the following discussion but only the handler execution time. Figure~\ref{fig:scaling} further shows the minimum, maximum, and the geometric mean for all handlers of Section~\ref{sec:apps}. We notice how, even for an 800 Gbit/s network and packet latencies of 3 us, the required packet buffer capacity is only 300 KiB. Currently, \name{} provides a 4 MiB packet buffer that would be enough to sustain even higher line rate or packet latencies. Moreover, we need to take into account that handlers should terminate within a threshold in order to not bottleneck the incoming data flow, constraining the packet buffer size. We plot the handler critical times (HCT) that are thresholds after which handlers become bottlenecks for given combination of packet size, line rate, and core count. HCTs are computed by scaling the number of cores together with the line rate. We show HCTs for 256 B, 1024 B, and 4096 B packets.
In Section~\ref{sec:hw-synthesis}, we show how memories take a large portion of both area (87\%) and power (52\%). Since the memories provided in this configuration are partially independent from the network bandwidth (i.e., handler memory) and partially over-provisioned (i.e., packet buffer), we expect area and power to remain stable while scaling \name{} to sustain higher line rates.

The modular organization of \name{} allows to scale up the number of processing clusters (\textbf{S8}) for, e.g., enabling higher workloads (i.e., longer handlers) without becoming a bottleneck. However, to increase the sustained network bandwidth, we would either need to re-balance the system in order to feed the processing clusters at line rate (e.g., to sustain 1 Tbit/s, we should adopt 1024-bit data paths), or have multiple \name accelerators with an additional scheduling level. 

In the near future, we plan to investigate a possible integration of Snitch~\cite{zaruba2020snitch} cores and clusters in \name{}. With simpler RISC-V cores, a more flexible cluster architecture, and virtual memory support, we believe this integration can further improve area and power efficiency, and increase the flexibility of the proposed in-network compute accelerator. Additionally, we are following a line of research aimed at evaluating the costs and benefits of having \name{} integrated into network switches, which would enable general packet processing deeper in the network.

%
%\begin{itemize}
%	\item Our packet buffer is enough (actually overestimated) also for higher line rates. Say also that we need to take into account the critical handler time, which should be the average handler time to not being bottleneck. 
%	\item The rest of the memory does not depend on the line rate, but on the application and number/type of handlers that are installed. 
%	\item Since both power and area are dominated by memories, we do not expect neither area nor power to go crazy high. 
%	\item To scale with line rate one can either increase the number of cores/cluster (problem is interconnect) or increase the number of clusters. Mention core Snitch and say that are smaller and we are considering it as future work (can pack more cores in the same area). 
%\end{itemize}

%\salvo{ault: 14 nm $\sim$~50.354 mm$^2$ die area; num of cores: 18; area/per core: $\sim$~2.80 mm$^2$}

%\salvo{ault TDP: 200 W https://en.wikipedia.org/wiki/Skylake\_(microarchitecture)}

%\salvo{pspin power?}

%\subsubsection{Throughput}
%\enlargethispage{\baselineskip}
%\clearpage
%\vspace{-0.7em}
\section{Related Work}
%\vspace{-0.4em}

%\enlargethispage{0.5\baselineskip}

% AM? Maybe too far away conceptually.
One of the oldest concepts related to \name is  Active
Messages~(AM)~\cite{activemessages}. However, in the AM model, messages are atomic and can be processed only once they are fully received. In sPIN, the processing happens at the packets level, leading to lower latencies and buffer requirements.

% Packet processing systems
sPIN is closely related to systems such as
P4~\cite{bosshart2014p4}, which allow users to define match-action rules on a
per-packet basis and are supported by switch architectures such as
AMT~\cite{bosshart2013forwarding}, FlexPipe~\cite{flexpipe}, and Cavium's
Xplaint. Those architectures target switches and work on packet headers, not
packet data. FlexNIC~\cite{flexnic} extends this idea by introducing modifiable memory and enabling fine-grained
steering of DMA streams at the receiver NIC.
These extensions can be used for,
e.g., partition the key-space of a key-value store and steer requests to specific cores.
However, the offloading of complex application-specific tasks (e.g., datatype processing~\cite{spin-ddt}) has not been demonstrated in this programming model.
In contrast, PsPIN allows offloading of arbitrary functions executed on
general-purpose processing cores with small hardware extensions to increase
throughput and reduce latency. 
\review{
PANIC~\mbox{\cite{lin2020panic}} is a recent work sharing many design principles of \name{}. The main difference is that \name{} allows applications to define handlers to be executed on incoming packets while PANIC enables them to express compositions of pre-offloaded tasks.}{E.5}

% Fully programmable NICs
Programmable NICs are not new. Quadrics QSNet employed them to
accelerate collectives~\cite{yu2004efficient} and to implement early versions
of Portals~\cite{pedretti2004nic}. Myrinet NICs~\cite{wagner2004nic}
allowed users to offload modules written in C to the specialized NIC cores.
Modern approaches to NIC offload~\mbox{\cite{sidler2020strom, eran2019nica}} requires network engineers to implement 
functionalities as FPGA modules, while \mbox{\name} uses easier to (re-)program RISC-V cores.
Differently from these approaches, sPIN enables user-space applications to define their own C/C++ packet handlers.

%\begin{itemize}
%\item Catapult \cite{catapult}
%\item Azure SmarNICs \cite{firestone2018azure}
%\end{itemize}

%\vspace{-1em}
\section{Conclusions}
\vspace{-0.4em}

Processing data in the network is a necessary step to scale applications along with the network speeds. 
This work 
defines principles and architectural characteristics of packet
processing engines, which are the next step after RDMA acceleration. We propose \name, a power and area efficient RISC-V based unit implementing the sPIN programming model, defining the interfaces for NIC integration. We evaluate \name, showing that it can process packets at up to 400 Gbit/s line rate and motivate our architectural choices with a performance study of a set of example handlers over different architectures. PsPIN is an open-source project and is available at:

\begin{center}
\url{https://github.com/spcl/pspin}
\end{center}
\section*{Acknowledgments}
This work has been partially funded by the EPIGRAM-HS project under grant agreement no.~801039, the European Union's H2020 Specific Grant Agreement for European Processor Initiative~(EPI) under grant agreement no.~826647, the Croatian-Swiss Research Programme under grant agreement no.~180625, and the European Project RED-SEA under grant agreement no.~955776.

%%%%%%% -- PAPER CONTENT ENDS -- %%%%%%%%
\balance
%%%%%%%%% -- BIB STYLE AND FILE -- %%%%%%%%
\bibliographystyle{plain}
\bibliography{refs}

\begin{thebibliography}{10}

\bibitem{broadcom}
{Broadcom Stingray SmartNIC}.
\newblock
  \url{https://www.broadcom.com/products/ethernet-connectivity/smartnic}.
\newblock Accessed: 2020-18-03.

\bibitem{bluefield}
{Mellanox BlueField SmartNIC}.
\newblock \url{https://www.mellanox.com/products/bluefield-overview}.
\newblock Accessed: 2020-18-03.

\bibitem{arm-area}
{Wikichip.org}: {Cortex-A53} - microarchitectures - {ARM}.
\newblock
  \url{https://en.wikichip.org/wiki/arm_holdings/microarchitectures/cortex-a53}.
\newblock Accessed: 2020-15-04.

\bibitem{ault-area}
{Wikichip.org}: {Skylake} (server) - microarchitectures - {Intel}.
\newblock
  \url{https://en.wikichip.org/wiki/intel/microarchitectures/skylake_(server)}.
\newblock Accessed: 2020-15-04.

\bibitem{ugnidmapp}
{XC Series GNI} and {DMAPP API} user guide.
\newblock
  \url{https://pubs.cray.com/content/S-2446/CLE%207.0.UP01/xctm-series-gni-and-dmapp-api-user-guide}.
\newblock Accessed: 2020-18-03.

\bibitem{alverson2012cray}
Bob Alverson, Edwin Froese, Larry Kaplan, and Duncan Roweth.
\newblock {Cray XC} series network.
\newblock {\em Cray Inc., White Paper WP-Aries 01-1112}, 2012.

\bibitem{alverson2010gemini}
Robert Alverson, Duncan Roweth, and Larry Kaplan.
\newblock The gemini system interconnect.
\newblock In {\em 2010 18th IEEE Symposium on High Performance Interconnects},
  pages 83--87. IEEE, 2010.

\bibitem{attig2011400}
Michael Attig and Gordon Brebner.
\newblock 400 {Gb/s} programmable packet parsing on a single {FPGA}.
\newblock In {\em 2011 ACM/IEEE Seventh Symposium on Architectures for
  Networking and Communications Systems}, pages 12--23. IEEE, 2011.

\bibitem{portals42}
B.~W Barrett, R.~Brightwell, S.~Hemmert, K.~Pedretti, K.~Wheeler, K.~Underwood,
  R.~Riesen, T.~Hoefler, A.~B Maccabe, and T.~Hudson.
\newblock The {Portals 4.2} network programming interface.
\newblock {\em Sandia National Laboratories, November 2012, Technical Report
  SAND2012-10087}, 2018.

\bibitem{barthels2017distributed}
Claude Barthels, Ingo M{\"u}ller, Timo Schneider, Gustavo Alonso, and Torsten
  Hoefler.
\newblock Distributed join algorithms on thousands of cores.
\newblock {\em Proceedings of the VLDB Endowment}, 10(5), 2017.

\bibitem{bedeir2010building}
Tarick Bedeir.
\newblock Building an {RDMA}-capable application with {IB Verbs}.
\newblock {\em Technical report, HPC Advisory Council}, 2010.

\bibitem{belshe2015hypertext}
Mike Belshe, Roberto Peon, and Martin Thomson.
\newblock Hypertext transfer protocol version 2 ({HTTP/2}), 2015.

\bibitem{bosshart2014p4}
Pat Bosshart, Dan Daly, Glen Gibb, Martin Izzard, Nick McKeown, Jennifer
  Rexford, Cole Schlesinger, Dan Talayco, Amin Vahdat, George Varghese, et~al.
\newblock {P4}: Programming protocol-independent packet processors.
\newblock {\em ACM SIGCOMM Computer Communication Review}, 44(3):87--95, 2014.

\bibitem{bosshart2013forwarding}
Pat Bosshart, Glen Gibb, Hun-Seok Kim, George Varghese, Nick McKeown, Martin
  Izzard, Fernando Mujica, and Mark Horowitz.
\newblock Forwarding metamorphosis: Fast programmable match-action processing
  in hardware for {SDN}.
\newblock {\em ACM SIGCOMM Computer Communication Review}, 43(4):99--110, 2013.

\bibitem{brunella2020hxdp}
Marco~Spaziani Brunella, Giacomo Belocchi, Marco Bonola, Salvatore Pontarelli,
  Giuseppe Siracusano, Giuseppe Bianchi, Aniello Cammarano, Alessandro Palumbo,
  Luca Petrucci, and Roberto Bifulco.
\newblock {hXDP}: Efficient software packet processing on {FPGA} {NICs}.
\newblock {\em arXiv preprint arXiv:2010.14145}, 2020.

\bibitem{buntinas2000fast}
Darius Buntinas, Dhabaleswar~K Panda, and Ponnuswamy Sadayappan.
\newblock Fast {NIC}-based barrier over {Myrinet/GM}.
\newblock In {\em Proceedings 15th International Parallel and Distributed
  Processing Symposium. IPDPS 2001}, pages 8--pp. IEEE, 2000.

\bibitem{ycsb}
Brian~F Cooper, Adam Silberstein, Erwin Tam, Raghu Ramakrishnan, and Russell
  Sears.
\newblock Benchmarking cloud serving systems with {YCSB}.
\newblock In {\em Proceedings of the 1st ACM symposium on Cloud computing},
  pages 143--154, 2010.

\bibitem{deri2007high}
Luca Deri.
\newblock High-speed dynamic packet filtering.
\newblock {\em Journal of Network and Systems Management}, 15(3):401--415,
  2007.

\bibitem{derradji2015bxi}
Sa{\"\i}d Derradji, Thibaut Palfer-Sollier, Jean-Pierre Panziera, Axel Poudes,
  and Fran{\c{c}}ois~Wellenreiter Atos.
\newblock The {BXI} interconnect architecture.
\newblock In {\em 2015 IEEE 23rd Annual Symposium on High-Performance
  Interconnects}, pages 18--25. IEEE, 2015.

\bibitem{spin-ddt}
Salvatore Di~Girolamo, Konstantin Taranov, Andreas Kurth, Michael Schaffner,
  Timo Schneider, Jakub Ber\'{a}nek, Maciej Besta, Luca Benini, Duncan Roweth,
  and Torsten Hoefler.
\newblock Network-accelerated non-contiguous memory transfers.
\newblock In {\em Proceedings of the International Conference for High
  Performance Computing, Networking, Storage and Analysis}, SC ’19, New York,
  NY, USA, 2019. Association for Computing Machinery.

\bibitem{eran2019nica}
Haggai Eran, Lior Zeno, Maroun Tork, Gabi Malka, and Mark Silberstein.
\newblock {NICA}: An infrastructure for inline acceleration of network
  applications.
\newblock In {\em 2019 {USENIX} Annual Technical Conference ({USENIX} {ATC}
  19)}, pages 345--362, 2019.

\bibitem{firestone2018azure}
Daniel Firestone, Andrew Putnam, Sambhrama Mundkur, Derek Chiou, Alireza
  Dabagh, Mike Andrewartha, Hari Angepat, Vivek Bhanu, Adrian Caulfield, Eric
  Chung, et~al.
\newblock {Azure} accelerated networking: {SmartNICs} in the public cloud.
\newblock In {\em 15th $USENIX$ Symposium on Networked Systems Design and
  Implementation NSDI 18)}, pages 51--66, 2018.

\bibitem{gautschi2017near}
Michael Gautschi, Pasquale~Davide Schiavone, Andreas Traber, Igor Loi, Antonio
  Pullini, Davide Rossi, Eric Flamand, Frank~K G{\"u}rkaynak, and Luca Benini.
\newblock Near-threshold {RISC-V} core with {DSP} extensions for scalable {IoT}
  endpoint devices.
\newblock {\em IEEE Transactions on Very Large Scale Integration (VLSI)
  Systems}, 25(10):2700--2713, 2017.

\bibitem{graham2016scalable}
Richard~L Graham, Devendar Bureddy, Pak Lui, Hal Rosenstock, Gilad Shainer, Gil
  Bloch, Dror Goldenerg, Mike Dubman, Sasha Kotchubievsky, Vladimir Koushnir,
  et~al.
\newblock Scalable hierarchical aggregation protocol ({SHArP}): a hardware
  architecture for efficient data reduction.
\newblock In {\em 2016 First International Workshop on Communication
  Optimizations in HPC (COMHPC)}, pages 1--10. IEEE, 2016.

\bibitem{connectx}
Richard~L Graham, Steve Poole, Pavel Shamis, Gil Bloch, Noam Bloch, Hillel
  Chapman, Michael Kagan, Ariel Shahar, Ishai Rabinovitz, and Gilad Shainer.
\newblock {ConnectX-2} {InfiniBand} management queues: First investigation of
  the new support for network offloaded collective operations.
\newblock In {\em 2010 10th IEEE/ACM International Conference on Cluster, Cloud
  and Grid Computing}, pages 53--62. IEEE, 2010.

\bibitem{gurkaynak2017multi}
Frank~K G{\"u}rkaynak, Robert Schilling, Michael Muehlberghuber, Francesco
  Conti, Stefan Mangard, and Luca Benini.
\newblock Multi-core data analytics {SoC} with a flexible {1.76 Gbit/s}
  {AES-XTS} cryptographic accelerator in {65 nm} {CMOS}.
\newblock In {\em Proceedings of the Fourth Workshop on Cryptography and
  Security in Computing Systems}, pages 19--24, 2017.

\bibitem{handy1998cache}
Jim Handy.
\newblock {\em The cache memory book}.
\newblock Morgan Kaufmann, 1998.

\bibitem{spin}
Torsten Hoefler, Salvatore Di~Girolamo, Konstantin Taranov, Ryan~E. Grant, and
  Ron Brightwell.
\newblock {sPIN}: High-performance streaming processing in the network.
\newblock In {\em Proceedings of the International Conference for High
  Performance Computing, Networking, Storage and Analysis}, SC '17, New York,
  NY, USA, 2017. Association for Computing Machinery.

\bibitem{hoefler2015remote}
Torsten Hoefler, James Dinan, Rajeev Thakur, Brian Barrett, Pavan Balaji,
  William Gropp, and Keith Underwood.
\newblock Remote memory access programming in {MPI-3}.
\newblock {\em ACM Transactions on Parallel Computing (TOPC)}, 2(2):1--26,
  2015.

\bibitem{infiniband2000infiniband}
{InfiniBand Trade Association} et~al.
\newblock {InfiniBand} architecture specification release 1.2.
\newblock {\em http://www.infinibandta.org}, 2000.

\bibitem{iyengar2018quic}
Jana Iyengar and Martin Thomson.
\newblock {QUIC}: A {UDP}-based multiplexed and secure transport.
\newblock {\em Internet Engineering Task Force, Internet-Draft
  draftietf-quic-transport-17}, 2018.

\bibitem{flexnic}
Antoine Kaufmann, SImon Peter, Naveen~Kr Sharma, Thomas Anderson, and Arvind
  Krishnamurthy.
\newblock High performance packet processing with {FlexNIC}.
\newblock In {\em ACM SIGARCH Computer Architecture News}, volume~44, pages
  67--81. ACM, 2016.

\bibitem{ebpf-netronome}
Jakub Kicinski and Nicolaas Viljoen.
\newblock {eBPF} hardware offload to {SmartNICs}: {clsbpf} and {XDP}.

\bibitem{kumar2012horizontal}
V~Pradeep Kumar and RV~Krishnaiah.
\newblock Horizontal aggregations in {SQL} to prepare data sets for data mining
  analysis.
\newblock {\em IOSR Journal of Computer Engineering (IOSRJCE)}, pages
  2278--0661, 2012.

\bibitem{kurth2020axi}
Andreas Kurth, Wolfgang R\"onninger, Thomas Benz, Matheus Cavalcante, Fabian
  Schuiki, Florian Zaruba, and Luca Benini.
\newblock An open-source platform for high-performance non-coherent on-chip
  communication.
\newblock {\em arXiv preprint arXiv:2009.05334}, 2020.

\bibitem{lin2020panic}
Jiaxin Lin, Kiran Patel, Brent~E Stephens, Anirudh Sivaraman, and Aditya
  Akella.
\newblock {PANIC}: A high-performance programmable {NIC} for multi-tenant
  networks.
\newblock In {\em 14th {USENIX} Symposium on Operating Systems Design and
  Implementation ({OSDI} 20)}, pages 243--259, 2020.

\bibitem{mair20164}
Hugh~T Mair, Gordon Gammie, Alice Wang, Rolf Lagerquist, CJ~Chung, Sumanth
  Gururajarao, Ping Kao, Anand Rajagopalan, Anirban Saha, Amit Jain, et~al.
\newblock A {20nm} {2.5 GHz} ultra-low-power tri-cluster {CPU} subsystem with
  adaptive power allocation for optimal mobile {SoC} performance.
\newblock In {\em 2016 IEEE International Solid-State Circuits Conference
  (ISSCC)}, pages 76--77. IEEE, 2016.

\bibitem{mpi-3.0}
{Message Passing Interface Forum}.
\newblock {MPI}: A message-passing interface standard version 3.0, 09 2012.
\newblock Chapter author for Collective Communication, Process Topologies, and
  One Sided Communications.

\bibitem{miano2018creating}
Sebastiano Miano, Matteo Bertrone, Fulvio Risso, Massimo Tumolo, and
  Mauricio~V{\'a}squez Bernal.
\newblock Creating complex network services with {eBPF}: Experience and lessons
  learned.
\newblock In {\em 2018 IEEE 19th International Conference on High Performance
  Switching and Routing (HPSR)}, pages 1--8. IEEE, 2018.

\bibitem{nickolls2008scalable}
John Nickolls, Ian Buck, Michael Garland, and Kevin Skadron.
\newblock Scalable parallel programming with {CUDA}.
\newblock {\em Queue}, 6(2):40--53, 2008.

\bibitem{flexpipe}
Recep Ozdag.
\newblock {Intel{\textregistered}} {Ethernet} switch {FM6000} series - software
  defined networking.
\newblock {\em See goo.gl/AnvOvX}, 5, 2012.

\bibitem{pedretti2004nic}
Ron Brightwell Kevin~T Pedretti and Ron Brightwell.
\newblock A {NIC}-offload implementation of {Portals} for {Quadrics QsNet}.
\newblock In {\em Fifth LCI International Conference on Linux Clusters}, 2004.

\bibitem{petrini2001hardware}
Fabrizio Petrini, Salvador Coll, Eitan Frachtenberg, and Adolfy Hoisie.
\newblock Hardware-and software-based collective communication on the
  {Quadrics} network.
\newblock In {\em Proceedings IEEE International Symposium on Network Computing
  and Applications. NCA 2001}, pages 24--35. IEEE, 2001.

\bibitem{pontarelli2019flowblaze}
Salvatore Pontarelli, Roberto Bifulco, Marco Bonola, Carmelo Cascone, Marco
  Spaziani, Valerio Bruschi, Davide Sanvito, Giuseppe Siracusano, Antonio
  Capone, Michio Honda, et~al.
\newblock {Flowblaze}: Stateful packet processing in hardware.
\newblock In {\em 16th {USENIX} Symposium on Networked Systems Design and
  Implementation ({NSDI} 19)}, pages 531--548, 2019.

\bibitem{pyo201523}
Jungyul Pyo, Youngmin Shin, Hoi-Jin Lee, Sung-il Bae, Min-su Kim, Kwangil Kim,
  Ken Shin, Yohan Kwon, Heungchul Oh, Jaeyoung Lim, et~al.
\newblock A {20nm} {high-K} metal-gate heterogeneous {64b} quad-core {CPUs} and
  hexa-core {GPU} for high-performance and energy-efficient mobile application
  processor.
\newblock In {\em 2015 IEEE International Solid-State Circuits
  Conference-(ISSCC) Digest of Technical Papers}, pages 1--3. IEEE, 2015.

\bibitem{dpdk}
R~Rajesh, Kannan~Babu Ramia, and Muralidhar Kulkarni.
\newblock Integration of {LwIP} stack over {Intel} {DPDK} for high throughput
  packet delivery to applications.
\newblock In {\em 2014 Fifth International Symposium on Electronic System
  Design}, pages 130--134. IEEE, 2014.

\bibitem{ramakrishnan2001addition}
Kadangode Ramakrishnan, Sally Floyd, David Black, et~al.
\newblock The addition of explicit congestion notification ({ECN}) to {IP}.
\newblock 2001.

\bibitem{pulp}
Davide Rossi, Francesco Conti, Andrea Marongiu, Antonio Pullini, Igor Loi,
  Michael Gautschi, Giuseppe Tagliavini, Alessandro Capotondi, Philippe
  Flatresse, and Luca Benini.
\newblock {PULP}: A parallel ultra low power platform for next generation {IoT}
  applications.
\newblock In {\em 2015 IEEE Hot Chips 27 Symposium (HCS)}, pages 1--39. IEEE,
  2015.

\bibitem{inca}
Whit Schonbein, Ryan~E Grant, Matthew~GF Dosanjh, and Dorian Arnold.
\newblock {INCA}: in-network compute assistance.
\newblock In {\em Proceedings of the International Conference for High
  Performance Computing, Networking, Storage and Analysis}, pages 1--13, 2019.

\bibitem{sensi2020indepth}
Daniele~De Sensi, Salvatore~Di Girolamo, Kim~H. McMahon, Duncan Roweth, and
  Torsten Hoefler.
\newblock An in-depth analysis of the {Slingshot} interconnect, 2020.

\bibitem{sidler2020strom}
David Sidler, Zeke Wang, Monica Chiosa, Amit Kulkarni, and Gustavo Alonso.
\newblock {StRoM}: smart remote memory.
\newblock In {\em Proceedings of the Fifteenth European Conference on Computer
  Systems}, pages 1--16, 2020.

\bibitem{stone2010opencl}
John~E Stone, David Gohara, and Guochun Shi.
\newblock {OpenCL}: A parallel programming standard for heterogeneous computing
  systems.
\newblock {\em Computing in science \& engineering}, 12(3):66--73, 2010.

\bibitem{subramanian2012remote}
Viswanath Subramanian, Michael~R Krause, and Ramesh VelurEunni.
\newblock Remote direct memory access ({RDMA}) completion, August~14 2012.
\newblock US Patent 8,244,825.

\bibitem{terpstra2010collecting}
Dan Terpstra, Heike Jagode, Haihang You, and Jack Dongarra.
\newblock Collecting performance data with {PAPI-C}.
\newblock In {\em Tools for High Performance Computing 2009}, pages 157--173.
  Springer, 2010.

\bibitem{traber2016pulpino}
Andreas Traber, Florian Zaruba, Sven Stucki, Antonio Pullini, Germain Haugou,
  Eric Flamand, Frank~K Gurkaynak, and Luca Benini.
\newblock {PULPino}: A small single-core {RISC-V} {SoC}.
\newblock In {\em 3rd RISCV Workshop}, 2016.

\bibitem{activemessages}
Thorsten Von~Eicken, David~E Culler, Seth~Copen Goldstein, and Klaus~Erik
  Schauser.
\newblock Active messages: a mechanism for integrated communication and
  computation.
\newblock {\em ACM SIGARCH Computer Architecture News}, 20(2):256--266, 1992.

\bibitem{wagner2004nic}
Adam Wagner, Hyun-Wook Jin, Dhabaleswar~K Panda, and Rolf Riesen.
\newblock {NIC}-based offload of dynamic user-defined modules for {Myrinet}
  clusters.
\newblock In {\em 2004 IEEE International Conference on Cluster Computing (IEEE
  Cat. No. 04EX935)}, pages 205--214. IEEE, 2004.

\bibitem{waterman2016risc}
Andrew Waterman, Yunsup Lee, Rimas Avizienis, David~A Patterson, and Krste
  Asanovi{\'c}.
\newblock The {RISC-V} instruction set manual volume {II}: Privileged
  architecture version 1.9.
\newblock {\em EECS Department, University of California, Berkeley, Tech. Rep.
  UCB/EECS-2016-129}, 2016.

\bibitem{waterman2014risc}
Andrew Waterman, Yunsup Lee, David Patterson, Krste Asanovic, and Volume I~User
  level Isa.
\newblock The {RISC-V} instruction set manual.
\newblock {\em Volume I: User-Level ISA', version}, 2, 2014.

\bibitem{yu2004efficient}
Weikuan Yu, Darius Buntinas, Richard~L Graham, and Dhabaleswar~K Panda.
\newblock Efficient and scalable barrier over {Quadrics} and {Myrinet} with a
  new {NIC}-based collective message passing protocol.
\newblock In {\em 18th International Parallel and Distributed Processing
  Symposium, 2004. Proceedings.}, page 182. IEEE, 2004.

\bibitem{zaruba2020snitch}
Florian Zaruba, Fabian Schuiki, Torsten Hoefler, and Luca Benini.
\newblock {Snitch}: A tiny pseudo dual-issue processor for area and energy
  efficient execution of floating-point intensive workloads.
\newblock {\em IEEE Transactions on Computers}, 2020.

\end{thebibliography}
%%%%%%%%%%%%%%%%%%%%%%%%%%%%%%%%%%%%

\end{document}